%% file: noncoop.tex
\documentclass[11pt,letterpaper]{article}
\usepackage[utf8]{inputenc}
\usepackage{amsmath}
\usepackage{amssymb}
\usepackage{amsfonts}
\usepackage{amsthm}
\usepackage[usenames,dvipsnames,svgnames,table]{xcolor}
\usepackage{graphicx}
\usepackage{color}
\usepackage{hyperref}
\usepackage{cite}
\usepackage{tikz}
\usepackage{xspace}
\usepackage{needspace}
\usepackage{url}
\usepackage{subcaption}
\usepackage[margin=1in]{geometry}

\newtheorem{theorem}{Theorem}[section]

\newtheorem{proposition}{Proposition}
\theoremstyle{definition}
\newtheorem{definition}[theorem]{Definition}

\newtheorem{openproblem}{Open Problem}
\newtheorem{algorithm}[theorem]{Algorithm}
\let\oldalgorithm\algorithm
\renewcommand{\algorithm}{\oldalgorithm\normalfont}

\newcommand{\termasm}[1]{\mathcal{A}_{\Box}[{#1}]}
\newcommand{\prodasm}[1]{\mathcal{A}[{#1}]}

\newcommand{\dom}[1]{{\rm dom}(#1)}

\newcommand{\vect}{\overrightarrow}
\newcommand{\resp}{respectively\xspace}
\newcommand\ignore[1]{}

\makeatletter
\newtheorem*{rep@theorem}{\rep@title}
\newcommand{\newreptheorem}[2]{%
\newenvironment{rep#1}[1]{%
 \def\rep@title{#2 \ref{##1}}%
 \begin{rep@theorem}}%
 {\end{rep@theorem}}}
\makeatother
\newreptheorem{theorem}{Theorem}

\newcommand\No{\mathrm{N}}
\newcommand\So{\mathrm{S}}
\newcommand\We{\mathrm{W}}
\newcommand\Ea{\mathrm{E}}

\newcommand\letexp{baggins expression\xspace}
\newcommand\Letexp{Baggins expression\xspace}

\title{Non-cooperative algorithms in self-assembly}
\author{Pierre-Etienne Meunier\thanks{Aix Marseille Université, CNRS, LIF UMR 7279, 13288, Marseille, France, \protect\url{pierre-etienne.meunier@lif.univ-mrs.fr}. Supported in part by National Science Foundation Grant CCF-1219274.}}
\date{}

\begin{document}
\maketitle

\begin{abstract}

  We show the first non-trivial positive algorithmic results (i.e. programs
  whose output is larger than their size), in a model of self-assembly that has
  so far resisted many attempts of formal analysis or programming: the planar
  non-cooperative variant of Winfree's abstract Tile Assembly Model.

  This model has been the center of several open problems and conjectures in the
  last fifteen years, and the first fully general results on its computational
  power were only proven recently (SODA 2014). These results, as well as ours,
  exemplify the intricate connections between computation and geometry that can
  occur in self-assembly.

  In this model, tiles can stick to an existing assembly as soon as one of their
  sides matches the existing assembly. This feature contrasts with the general
  \emph{cooperative} model, where it can be required that tiles match on
  \emph{several} of their sides in order to bind.

  In order to describe our algorithms, we also introduce a generalization of
  regular expressions called \emph{\letexp}. Finally, we compare this model to
  other automata-theoretic models.

\end{abstract}

\section{Introduction}

Self-assembly is the process by which unorganized atomic components coalesce
into complex shapes and structures in an unsupervised way. This kind of
processes is ubiquitous in nature, and in particular in the complex molecular
components of life. In recent years, its study has yielded a growing number of
impressive experimental realizations, ranging from regular
arrays~\cite{WinLiuWenSee98} to fractal
structures~\cite{RoPaWi04,FujHarParWinMur07}, smiling
faces~\cite{RothOrigami,wei2012complex}, DNA tweezers~\cite{yurke2000dna}, logic
circuits~\cite{seelig2006enzyme,qian2011scaling}, neural
networks~\cite{qian2011neural}, and molecular robots\cite{DNARobotNature2010}.

Potential future applications range from more efficient, cheaper
computational units to interactions with natural biological processes, both for
medical diagnosis and treatment, and a better understanding of evolution and
development.

Realizing that \emph{programming} these processes is the keystone of atomically
precise molecular engineering, Winfree introduced in 1998 the abstract Tile
Assembly Model~\cite{Winf98} to program assemblies using the components built by
Seeman~\cite{Seem82} using DNA. This model is similar to Wang
tilings~\cite{Wang61}, essentially augmented with a mechanism for sequential
growth, and thus allowing mismatches between adjacent tiles.  More precisely, in
the abstract Tile Assembly Model, we consider square tiles from a finite set of
\emph{types}, with \emph{colors} and integer \emph{glue strengths} on each
side. The assembly starts from a single ``seed'' tile, and proceeds by adding
one tile at a time, asynchronously and nondeterministically.  At each step, a
tile can stick to the current assembly if the glue strengths, on its sides whose
colors match the current assembly, sum up to at least a parameter of the model
called the \emph{temperature} $\tau=1,2,3\ldots$

In the present work, we are mostly interested in the case of temperature 1
self-assembly, also called non-cooperative self-assembly.
In the abstract Tile
Assembly Model, when the temperature increases, fewer assemblies are possible,
allowing more control over producible assemblies: for instance, cooperative
self-assembly (i.e. at temperature at least 2) is able to simulate arbitrary
Turing machines \cite{Winf98,RotWin00,jCCSA}, and produce arbitrary connected
shapes with a number of tile types within a log factor of their Kolmogorov
complexity \cite{SolWin07}.  More surprisingly, this model has even been shown
intrinsically universal \cite{IUSA}, meaning that there is a single tileset
capable of simulating arbitrary tile assembly systems, modulo rescaling, even
with a single tile type \cite{Demaine-2012}.

Despite its apparent simplicity, the non-cooperative model is far from being well
understood, and not known to be capable of general Turing computation. However, this
is a fundamental and ubiquitous form of growth in nature, as many systems, from
plants to mycelium to percolation processes, exhibit this kind of behavior by growing
and branching tips.

In one of the first studies on self-assembly \cite{RotWin00}, Rothemund and
Winfree conjectured it to be less powerful than cooperative self-assembly. The
first fully general separation result, without unproven hypotheses, was only
proven recently~\cite{Meunier-2014}, in the context of intrinsic
universality~\cite{USA,IUSA,2HAMIU}. Before that, several results had shown
separations between particular cases of the model \cite{Doty-2011,Reif-2012,
  Manuch-2010}, and general self-assembly.

One of the most puzzling results on this model is its capability to simulate
Turing machines in the three dimensional generalization of the
model~\cite{Cook-2011}, whereas in one dimension, it is equivalent to finite
automata.

\subsection{Main results}

Here, we present the first efficient constructions in the fully general planar
noncooperative model. The generally accepted definition of an ``efficient
program'', in this context, is a program whose output is larger than its size.
Of course, a simple first result on this model shows that arbitrary shapes can
be built with a number of tile types equal to the number of tiles in the shape,
or (for simpler shapes) equal to the Manhattan diameter of the
shape~\cite{RotWin00}.

Surprisingly, our results show that there are tile assembly systems whose
terminal assemblies are all larger (in Manhattan diameter) than their number of
tile types.  Although a number of terms have not been defined yet, we briefly
introduce our two main constructions. The first construction can be proven
easily by hand; we will demonstrate it first in Section \ref{sect:efficient},
and then generalize it in Section \ref{sect:general}, to get the following
theorem:

\begin{reptheorem}{thm:general}
For all integer $n$, there is a tile assembly system
$\mathcal{T}_n=(T_n,\sigma_n,1)$ such that $|T_n|=n$, and for all terminal
assembly $a\in\termasm{\mathcal{T}_n}$, $a$ is finite and of height
$2n+o(n)$.
\end{reptheorem}

Intuitively, this construction works by preventing subpaths starting and ending
with the same tile type to repeat completely. However, it does not address the
possibility that some paths be efficient by repeating a subpath several times,
before being blocked. Since these partial pumping have been a major puzzle of
the field, we provide a second efficient construction allowing it.

However, its proof is significantly more complicated, and a generalized form of
our construction does not seem easy. In Section \ref{sect:partially}, we present
the computer-aided proof of its efficiency, that we have needed due to the size
at which the first ``savings'' of tile types are seen. Computer-aided proofs are
of growing importance in computer science and mathematics, as exemplified by its
latest developments in complexity theory, also in the context of tile
assembly~\cite{Kopecki-2014}.

However, our case here is significantly simpler, since this construction could
be verified by hand, probably within a few hours:

\begin{reptheorem}{thm:partially}
  There is a tile assembly system $\mathcal{T}=(T,\sigma,1)$ such that
  $|\dom\sigma|=1$, and all terminal assemblies of $\mathcal{T}$ contain a path $P$
  of Manhattan diameter strictly larger than $|T|+1$, that is partially pumped, i.e.
  parts of $P$ are consecutive repetitions of one of its subpath.
\end{reptheorem}

Finally, we compare this model, and several generalizations of it, to various
models of automata: finite automata, tree automata, and pushdown automata. These
models are explained in Section \ref{sect:comp}, and the comparison is
summarized on Figure \ref{fig:comp}.

\subsection{Key technical ideas and methods}

A major challenge, when studying non-cooperative self-assembly, is to overcome
the intuition given by the one-dimensional case (which is equivalent to finite
automata), that any repetition of a tile type may allow to ``pump'' an assembly.
Indeed, an easy observation shows that assemblies formed at temperature 1 are
nothing more than a collection of paths growing from the seed: if a tile type is
ever repeated along a path, it is tempting to try to repeat the subpath between
these repetitions.

However, geometry makes things more complex. First, there are simple
counter-examples to this pumping idea. Moreover, paths could first lay
``blocking parts'' out, and then come back and branch to check which type of
blocker has been formed; this is for instance the primary mechanism used by the
simulation of Turing machines in 3d shown in \cite{Cook-2011}. However, their
construction ``fakes cooperation'' by laying a blocker out for all alternatives
but one.

On the other hand, recent (unpublished) progresses tend to show that this kind
of ``bit reading'' gadgets is not possible in two dimensions. This model thus
asks a different question: can you write efficient programs without the ability
to read your workspace?

Our results show that this is possible, at least to some extent. They do so by
carefully considering the fact that paths that are monotonic in one dimension
are pumpable; therefore, we must build ``caves'', i.e. subpaths that are
non-monotonic in both dimensions. However, since these are more expensive to
build than straight paths, we also need to reuse these extra tile types several
times, either by making these subpaths self-blocking (in Section
\ref{sect:general}), and branching before the blocking, or by allowing
\emph{some} pumping (in Section \ref{sect:partially}) before blocking it.

These results are quite puzzling and counter-intuitive; however, they do not
seem to make Turing computation possible. Therefore, a natural question is the
exact power of this model, that depends strongly on geometry, and that
no other ``classical'' model seems to capture, as shown in Section \ref{sect:comp}.

\section{Definitions and preliminaries}
\label{definitions}

We begin by defining the abstract tile assembly model, in a slightly more
general framework than usually. Let $G$ be a group with $n$ generators
$\vect{i_0},\vect{i_1},\ldots,\vect{i_{n-1}}$, and arbitrary relators. We will
use $G$ to define the geometric space: for instance, $\mathbb{Z}^2$ has two generators
$\vect{i_0}=(1,0)$ and $\vect{i_1}=(0,1)$, and one relation $\vect{i_0}\vect{i_1}=\vect{i_1}\vect{i_0}$.

A \emph{tile type} is a unit square with $2n$ sides, each consisting of a glue
\emph{label} and a nonnegative integer \emph{strength}. In the most common case
where $n=2$, we call a tile's sides north, east, south, and west, respectively,
according to the following picture:
\begin{center}
\begin{tikzpicture}[scale=0.8]
\draw(-0.75,-0.75)rectangle(0.75,0.75);
\draw[->](0,0)--(1.5,0)node[anchor=west]{$\vect{i_0}$ (East)};
\draw[->](0,0)--(0,1.5)node[anchor=south]{$\vect{i_1}$ (North)};
\draw[->](0,0)--(-1.5,0)node[anchor=east]{$-\vect{i_0}$ (West)};
\draw[->](0,0)--(0,-1.5)node[anchor=north]{$-\vect{i_1}$ (South)};
\end{tikzpicture}
\end{center}

Also, we write these directions $\No$, $\Ea$, $\So$ and $\We$, respectively.
When there is no ambiguity, we also write $\No(t)$, $\Ea(t)$, $\So(t)$ and $\We(t)$,
to mean the north, east, south and west glue of tile type $t$, respectively.
Moreover, for each direction $d$, we write $-d$ its opposite
direction.  We assume a finite set $T$ of tile types, but an infinite supply of
copies of each type. An \emph{assembly} is a positioning of the tiles on the
Cayley graph of $G$, %
that is, a partial function $\alpha:G\dashrightarrow T$.  To simplify the
notations, we will assume $G=\mathbb{Z}^2$ throughout the paper, unless
explicitly mentioned.

In this context, we say that two elements $g_0,g_1\in G$ are \emph{adjacent} if
$g_1=g_0+\vect{i_k}$ (\resp $g_1=g_0-\vect{i_k}$) for some generator
$\vect{i_k}$. In this case, their \emph{abutting side} is the $\vect{i_k}$
side  (\resp the $-\vect{i_k}$ side) of $g_0$, and the $-\vect{i_k}$ side
(\resp the $\vect{i_k}$ side) of $g_1$.

We say that two tiles in an assembly \emph{interact}, or are \emph{stably
  attached}, if the glue labels on their abutting side are equal, and have
positive strength.  An assembly $\alpha$ induces a weighted \emph{binding graph}
$G_\alpha=(V_\alpha,E_\alpha)$, where $V_\alpha=\dom{\alpha}$ (the domain of
$\alpha$), and there is an edge $(a,b)\in E_\alpha$ if and only if $a$ and $b$
interact, and this edge is weighted by the glue strength of that interaction.
The assembly is said to be $\tau$-stable if any cut of $G_\alpha$ has weight at
least $\tau$.

A \emph{tile assembly system} is a triple $\mathcal{T}=(T,\sigma,\tau)$, where
$T$ is a finite tile set, $\sigma$ is called the \emph{seed}, and $\tau$ is the
\emph{temperature}. Throughout this paper, we will always have $\tau=1$, and
$\sigma$ will always be an assembly with exactly one tile. Therefore, we can
make the simplifying assumption that all glues have strength one without
changing the behavior of the model.

Given two $\tau$-stable assemblies $\alpha$ and $\beta$, we say that $\alpha$ is a
\emph{subassembly} of $\beta$, and write $\alpha\sqsubseteq\beta$, if
$\dom{\alpha}\subseteq \dom{\beta}$ and for all $p\in \dom{\alpha}$,
$\alpha(p)=\beta(p)$.
We also write
$\alpha\rightarrow_1^{\mathcal{T}}\beta$ if we can get $\beta$ from
$\alpha$ by the binding of a single tile, that is,
if $\alpha\sqsubseteq \beta$ and $|\dom{\beta}\setminus\dom{\alpha}|=1$.  We say that $\gamma$ is
\emph{producible} from $\alpha$, and write
$\alpha\rightarrow^{\mathcal{T}}\gamma$ if there is a (possibly empty)
sequence $\alpha=\alpha_1,\ldots,\alpha_n=\gamma$ such that
$\alpha_1\rightarrow_1^{\mathcal{T}}\ldots\rightarrow_1^{\mathcal{T}}\alpha_n$.

A sequence of $k\in\mathbb{Z}^+ \cup \{\infty\}$ assemblies
$\alpha_0,\alpha_1,\ldots$ over $\mathcal{A}^T$ is a
\emph{$\mathcal{T}$-assembly sequence} if, for all $1 \leq i < k$,
$\alpha_{i-1} \to_1^\mathcal{T} \alpha_{i}$.

The set of \emph{productions} of a tile assembly system $\mathcal{T}=(T,\sigma,\tau)$,
written $\prodasm{\mathcal{T}}$, is the set of all assemblies producible
from $\sigma$. An assembly $\alpha$ is called \emph{terminal} if there
is no $\beta$ such that $\alpha\rightarrow_1^{\mathcal{T}}\beta$. The
set of terminal assemblies is written $\termasm{\mathcal{T}}$.

The \emph{Manhattan distance} $\|\vect{AB}\|_1$ between two points $A=(x_A,y_A)$
and $B=(x_B,y_B)$ is $\|\vect{AB}\|_1=|x_A-x_B| + |y_A-y_B|$. The
  \emph{Manhattan diameter} of a connected assembly is the maximal Manhattan
  distance between two points in the assembly.
We write $(u_n)_{n\in\mathbb{N}}$ to mean ``the infinite sequence $u_0$,
$u_1$, $u_2$, $\ldots$''.

A \emph{regular tree grammar} $G=(S,N,\mathcal{F},R)$, according to
\cite{tata2007}, is given by an \emph{axiom} $S$, a set $N$ of \emph{nonterminal
  symbols}, a set $\mathcal{F}$ of \emph{terminal symbols}, and a set $R$ of
\emph{production rules} of the form $A\rightarrow\beta$ where $A$ is a
nonterminal and $\beta$ is a tree whose nodes are labeled by elements of
$\mathcal{F}\cup N$. Moreover, it is required that $\mathcal{F}\cap
N=\emptyset$.  In this work, we write trees as ``nested function
applications'': for instance, $f(x,g(y,z))$ is the following tree:

\begin{center}
\begin{tikzpicture}

\draw(0,2)--(-0.5,1)node[fill=white]{$x$};
\draw(0,2)--(0.5,1)--(0,0)node[fill=white]{$y$};
\draw(0.5,1)--(1,0)node[fill=white]{$z$};
\draw(0.5,1)node[fill=white]{$g$};
\draw(0,2)node[fill=white]{$f$};

\end{tikzpicture}
\end{center}

The classical example of a regular tree grammar is the grammar of lists of
integers, with one axiom $List$, non-terminals $List$ and $Nat$, terminals
$0$, $nil$, $s()$ and $cons(,)$, and the following rules:

\begin{eqnarray*}
List&\rightarrow&nil\\
List&\rightarrow&cons(Nat,List)\\
Nat&\rightarrow&0\\
Nat&\rightarrow&s(Nat)
\end{eqnarray*}

\section{Efficient algorithms}
\label{sect:algos}

In this section, we show the main ideas of our efficient tileset. In order to describe
them unambiguously, we use two different tools: figures showing the complete tileset
and seed on the one hand, and programs written in a generalization of regular expressions
called \emph{\letexp{}s}. An implementation of these expressions using a ``sublanguage''
of Haskell (i.e. a monad) is available
at \href{http://hackage.haskell.org/package/Baggins}{\tt http://hackage.haskell.org/package/Baggins}.

Moreover, all the constructions of this paper were generated in this language, and their
source code is available on
the self-assembly wiki\footnote{
\href{http://self-assembly.net/wiki/index.php?title=Baggins-expressions}{\tt http://self-assembly.net/wiki/index.php?title=Baggins-expressions}
}.
\subsection{\Letexp{}s}
\renewcommand\i[1]{\mathtt{#1}}
\newcommand\e[1]{\mathrm{#1}}
\newcommand\id[1]{\mathit{#1}}

A program in this language is an $\e{expr}$, where $\e{expr}$ is defined by the
following grammar (where an $\id{identifier}$ is a name):

\begin{eqnarray*}
\e{expr}&:=&\e{atom}\ |\ \e{let}\ |\ \e{bind}\ |\ \e{from}\ |\ \e{expr}\ \i{;}\ \e{expr}\\
\e{atom}&:=&\i{moveN}\ |\ \i{moveE}\ |\ \i{moveS}\ |\ \i{moveW}\\
\e{let}&:=&\i{let}\ \id{identifier}\\
\e{bind}&:=&\i{bind}\ [\ \i N\ |\ \i E\ |\ \i S\ |\ \i W\ ]\ \id{identifier}\\
\e{from}&:=&\i{from}\ \id{identifier}
\end{eqnarray*}

\begin{definition}
  \label{def:semantics}
  Let $e$ be a \letexp. Let $\beta$ the set of its
  identifiers. We define the unique tileset described by $e$ by induction on
  $e$:

  Let $T_0$ be a tileset consisting of a unique tile type $\sigma_0$,
  $C_0=\sigma_0$ and $\alpha_0$ is the function defined nowhere. Then, for all
  $i\in\{0,1,\ldots,|e|-1\}$:
  \begin{itemize}

  \item If $e_i=\i{moveN}$, $C_{i+1}=(g_\No,g_\Ea,\No(C_i),g_\We)$, and
    $T_{i+1}=T_i\cup\{C_{i+1}\}$, where
    $g_\No$, $g_\Ea$, $g_\We$ are all new glues, not appearing on any tile
    of $T_i$. Moreover, let $\alpha_{i+1}=\alpha$.
  \item If $e_i=\i{moveS}$, $C_{i+1}=(\So(C_i),g_\Ea,g_\So,g_\We)$, and
    $T_{i+1}=T_i\cup\{C_{i+1}\}$, where
    $g_\So$, $g_\Ea$, $g_\We$ are all new glues, not appearing on any tile
    of $T_i$. Moreover, let $\alpha_{i+1}=\alpha$.
  \item If $e_i=\i{moveE}$, $C_{i+1}=(g_\No,\We(C_i),g_\So,g_\We)$, and
    $T_{i+1}=T_i\cup\{C_{i+1}\}$, where
    $g_\No$, $g_\So$, $g_\We$ are all new glues, not appearing on any tile
    of $T_i$. Moreover, let $\alpha_{i+1}=\alpha$.
  \item If $e_i=\i{moveW}$, $C_{i+1}=(g_\No,g_\Ea,g_\So,\Ea(C_i))$, and
    $T_{i+1}=T_i\cup\{C_{i+1}\}$, where
    $g_\No$, $g_\Ea$, $g_\So$ are all new glues, not appearing on any tile
    of $T_i$. Moreover, let $\alpha_{i+1}=\alpha$.

  \item If $e_i=\i{let}\ x$, then let $\alpha_{i+1}$ be the function of domain
    $\dom{\alpha_i}\cup\{x\}$, such that for all
    $y\in\dom{\alpha_i}\setminus\{x\}$, $\alpha_{i+1}(y)=\alpha_i(y)$, and
    $\alpha_{i+1}(x)=C_i$.
  \item If $e_i=\i{bind}\ d\ x$, where $d\in\{\No,\So,\Ea,\We\}$ and
    $x\in\alpha_i$, then:
    \begin{itemize}
      \item $C_{i+1}=C_i$,
      \item $\alpha_{i+1}=\alpha_i$, and
      \item let $g$ be the glue on side $d$ of $C_i$, and $-g$ be the
        glue on side $-d$ of $\alpha_i(x)$. Then $T_{i+1}$ is $T_i$
        where all glues on sides $d$ and $-d$, that are equal to $g'$,
        are replaced with $g$.
      \end{itemize}
    \item If $e_i=\i{from}\ x$, where $x\in\alpha_i$, let
      $T_{i+1}=T_i$, $\alpha_{i+1}=\alpha_i$, and $C_i=\alpha_i(x)$.
  \end{itemize}
\end{definition}

\begin{theorem}

  Definition \ref{def:semantics} is ``sound and complete'', i.e. any \letexp describes exactly one tile assembly system, and any
  single-seeded tile assembly system can be described by a \letexp.

  \begin{proof}
    We prove the two properties independently:
    \begin{itemize}
    \item First remark that the construction of Definition \ref{def:semantics}
      defines a tileset and a seed non-ambiguously.

    \item Now, let $\mathcal{T}=(T,\sigma,1)$ be a temperature 1 tile assembly
      system with $|\dom\sigma|=1$. Start with $D=\{\sigma\}$. Then, for each
      tile $t\in T\setminus D$ that can bind to a tile $t_0\in D$ on side
      $d\in\{\No,\So,\Ea,\We\}$ of $t_0$, add $\i{from}\ t_0\ \i{move}d$ to
      $D$. Also, from any previously created tile $t_1$ that can bind to $t_0$,
      add $\i{from}\ t_0\ \i{bind}\ d\ t_1$ to $D$, if this binding has not been
      defined before, either directly or by operation $\i{from}\ t_1\ \i{bind}\
      (-d)\ t_1$ (and do nothing else).

      Clearly, this \letexp describes $\mathcal{T}$, by
      Definition \ref{def:semantics}.
    \end{itemize}
  \end{proof}
\end{theorem}

In order to make the examples in the appendix shorter and more intuitive, the
actual language used in our examples differs slightly from this grammar. However,
all its instructions can clearly be written using \letexp constructs.

\subsection{A first efficient algorithm}
\label{sect:efficient}
In this section, we call a tile assembly system $\mathcal{T}=(T,\sigma,1)$
\emph{efficient} if there is an integer $r$, such that the Manhattan diameter of
all the terminal assemblies of $\mathcal{T}$ is strictly larger than $|T|+|\dom\sigma|$,
and at most $r$.

A simple observation on paths, is that any path that is monotonic in one
dimension (i.e. the sequence $(y_{P_i})_{i}$ of its y-coordinates, or the
sequence $(x_{P_i})_i$ of its x-coordinates is monotonic), and repeats a tile
type, is pumpable.

Therefore, the main ingredient of efficient paths is non-monotonicity: we call a
\emph{vertical cave} (\resp \emph{horizontal cave}) a part of a path $P$ between
two indices $i$ and $j$, such that (1) $y_{P_i}=y_{P_j}$, (2) for all $k<i$,
$y_{P_k}\leq y_{P_i}$, and (3) for all $k\in\{i+1,i+2,\ldots,j-1\}$,
$y_{P_k}<y_{P_i}$.

Our first tile assembly system $\mathcal{T}_0$ is presented completely in
Appendix \ref{firsteff}, in the form of a \letexp.
We prove it now:

\begin{theorem}
\label{efficientpaths}
For all integer $n$, there is a tile assembly system
$\mathcal{T}_n=(T_n,\sigma_n,1)$ such that $|T_n|=n$, and for all terminal
assembly $a\in\termasm{\mathcal{T}_n}$, $a$ is finite and of height
$\frac{5(n+2)}{4}-23$.

\begin{proof}
Let $T_0$ be the set of tiles appearing on the lower right assembly of Figure
\ref{figure:efficient}, and $\sigma_0$ be the upper left assembly of that figure.

This tileset has 38 tile types, and its terminal assemblies are of height 27; it
is not efficient yet. But we will now add a number of new tile types to make it
efficient. First replace the following glues (zoom in on Figure
\ref{figure:efficient} to see these glue numbers, or see the large version in
Appendix~\ref{printable-eff3}):

\begin{itemize}
\item glue 6 by $(6,0)$ on the north, and $(6,n)$ on the south,
\item glue 14 by $(14,0)$ on the north, and $(14,n)$ on the south,
\item glue 24 by $(24,0)$ on the north, and $(24,n)$ on the south,
\item glue 26 by $(26,0)$ on the north, and $(26,n)$ on the south,
\end{itemize}

And then for all $i\in\{6,14,24,26\}$ and $j\in\{0,1,\ldots,n-1\}$, add a tile
type to $T$, with south glue $(i,j)$ and north glue $(i,j+1)$.  In total, we
have added $4n$ tile types, but the terminal assemblies of $T$ grow $5n$ higher.
See Figure \ref{figure:reallyefficient} for a larger example (saving tile type).

\begin{figure}[!ht]

\hspace{2cm}
\rotatebox{270}{
\begin{tikzpicture}
\input{eff1}
\end{tikzpicture}
}\hfill
\rotatebox{270}{
\begin{tikzpicture}
\input{eff2}
\end{tikzpicture}
}
\hspace{2cm}

\hspace{2cm}
\rotatebox{270}{
\begin{tikzpicture}
\input{eff0}
\end{tikzpicture}
}\hfill
\rotatebox{270}{
\begin{tikzpicture}
\input{eff3}
\end{tikzpicture}
}
\hspace{2cm}

\caption{Four successive stages of the construction: first the seed, then the
  main path grows, and finally, additional branches can also grow completely,
  along the main path.}
\label{figure:efficient}
\end{figure}
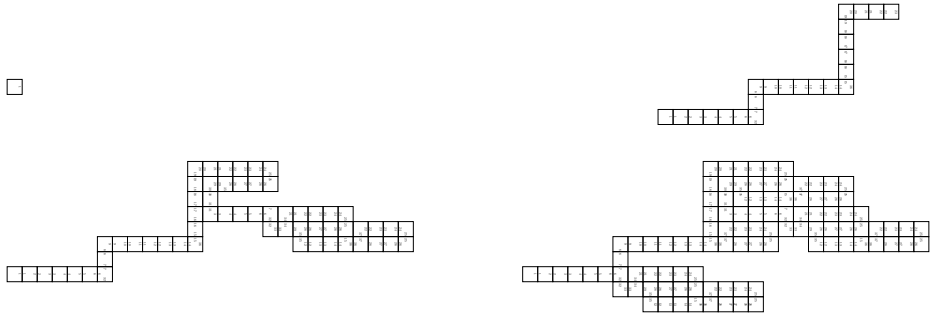

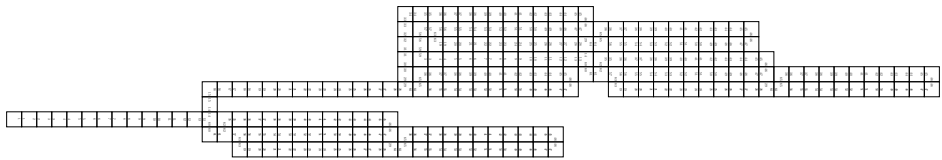
\begin{figure}[!ht]
\begin{center}

  \begin{tikzpicture}
    \draw[use as bounding box,draw=none](0,-2)rectangle(12.4,0);
    \begin{scope}[transform canvas={rotate=270}]
      \input{eff4}
    \end{scope}
  \end{tikzpicture}
\end{center}
\caption{An efficient tile assembly system, producing an assembly of width 112
  with 106 tile types. This terminal assembly grew from a seed containing only
  its leftmost tile.}
\label{figure:reallyefficient}
\end{figure}
\end{proof}
\end{theorem}

\subsection{A more general scheme}
\label{sect:general}

In the construction of Theorem \ref{efficientpaths}, repetitions of a tile type are done
at the expense of width of the assembly: indeed, in order to avoid collisions between
repeated paths, each repetition needs to be more and more narrow.
Generalizing this remark yields the following Theorem:

\begin{theorem}\label{thm:general}
For all integer $n$, there is a tile assembly system
$\mathcal{T}_n=(T_n,\sigma_n,1)$ such that $|T_n|=n$, and for all terminal
assembly $a\in\termasm{\mathcal{T}_n}$, $a$ is finite and of height
$2n+o(n)$.

\begin{proof}

  The idea is to repeat the construction of Theorem \ref{efficientpaths} more
  than a constant number of times. A single cave, of height $h$ (see Figure
  \ref{fig:constr}), will be reused $N$ times, and at each iteration
  $i\in\{0,1,\ldots,N\}$, grow to height $2h-i$.

  To do this, we use a sequence of assemblies as shown on Figure
  \ref{fig:constr}, with different widths $(w_n)_n$. The precise definition of
  this construction is given by the Haskell program in Appendix \ref{general},
  but the general idea is: grow some construction starting with tile type $t$,
  then use some modification of the initial cave as a blocker, and then reuse
  $t$.

  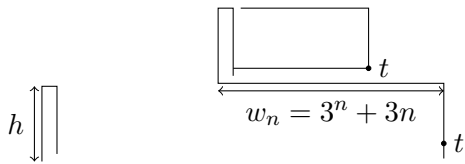
\begin{figure}[ht]
  \begin{center}
    \begin{tikzpicture}[scale=0.1]
      \draw(0,10)--(0,20)--(2,20)--(2,11);
      \draw[<->](-1,10)--(-1,20);
      \draw(-1,15)node[anchor=east]{$h$};
    \end{tikzpicture}\hspace{2cm}
    \begin{tikzpicture}[scale=0.1]
      \draw(30,0)--(30,10)--(0,10)--(0,20)--(2,20)--(2,11);
      \draw[fill=black](30,2)circle(0.3);
      \draw(30,2)node[anchor=west]{$t$};
      \draw(2,12)--(20,12)--(20,20)--(3,20);
      \draw[fill=black](20,12)circle(0.3);
      \draw(20,12)node[anchor=west]{$t$};
      \draw[<->](0,9)--(30,9);
      \draw(15,9)node[anchor=north]{$w_n=3^n+3n$};
    \end{tikzpicture}
  \end{center}
  \caption{The repeated part is shown on the left-hand side. The drawing on the
    right-hand side is a scheme of one step of the construction.}
  \label{fig:constr}
  \end{figure}

  Then, we stack these parts on top of each other: on the Figure \ref{fig:constr2}, the
  next assembly, drawn in dashed line, is of width $w_{n-1}=3^{n-1}+3(n-1)$. In order
  to avoid making a pumpable path, we do not grow the full initial cave each time, but
  a smaller and smaller suffix of it at each iteration.

  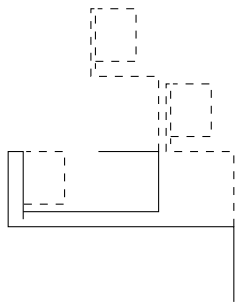
\begin{figure}[ht]
    \begin{center}
      \begin{tikzpicture}[scale=0.1]
        \draw(30,0)--(30,10)--(0,10)--(0,20)--(2,20)--(2,11);
        \draw(2,12)--(20,12)--(20,20)--(12,20);
        \begin{scope}[shift={(1.5,1)},dashed]
          \begin{scope}[xscale=0.3]
            \draw(2,12)--(20,12)--(20,19)--(3,19);
          \end{scope}
        \end{scope}

        \begin{scope}[shift={(21,10)},dashed]
          \begin{scope}[xscale=0.3]
            \draw(30,0)--(30,10)--(0,10)--(0,19)--(2,19)--(2,11);
            \draw(2,12)--(20,12)--(20,19)--(3,19);
          \end{scope}
        \end{scope}

        \begin{scope}[shift={(11,20)},dashed]
          \begin{scope}[xscale=0.3]
            \draw(30,0)--(30,10)--(0,10)--(0,19)--(2,19)--(2,11);
            \draw(2,12)--(20,12)--(20,19)--(3,19);
          \end{scope}
        \end{scope}
      \end{tikzpicture}
    \end{center}
    \caption{Two successive iterations.}
    \label{fig:constr2}
  \end{figure}
  Because of this choice of widths, successive assemblies cannot collide with
  each other, and different repetitions of the same assembly cannot collide with
  each other either.

  Let $h$ be the height of the initial cave.  For all integer $n$, the $n^{th}$
  repetition requires $w_n+2w_{n-1}\leq 2w_n$ new tiles horizontally, $h-n$
  tiles vertically, and grows to a height of $2(h-n)$. If we decide to repeat
  the construction $N=\log h$ times, we need
  $|T|=2\sum_{i=1}^N w_n + N h + O(N^2)$ tile types, i.e. $h\log h+O(h)$ tile types.

  Moreover, in this case, all terminal assemblies will have height
  $2h\log h+O(N^2)$, which is $2|T|+o(|T|)$.

  The \letexp for the exact construction is in Appendix
  \ref{general}.

\end{proof}
\end{theorem}

\subsection{Partially pumpable paths}
\label{sect:partially}
The constructions of Sections \ref{sect:efficient} and \ref{sect:general} are
efficient by repeating smaller and smaller parts of an assembly, while ensuring
that the assembly does not become pumpable. The other way of building efficient
paths is by letting them become pumpable for some time, after building
structures that block these repetitions. However, blocking these parts is
provably expensive, and the same kind of repeated blocking structure, similar to
those of Sections \ref{sect:efficient} and \ref{sect:general}, must be used to
``save'' tile types. However, this construction is intended as a proof that
allowing some pumping still does not forbid the existence of efficient tilesets.

\begin{theorem}\label{thm:partially}
  There is a tile assembly system $\mathcal{T}=(T,\sigma,1)$ such that
  $|\dom\sigma|=1$, and all terminal assemblies of $\mathcal{T}$ contain a path $P$
  of Manhattan diameter strictly larger than $|T|+1$, that is partially pumped, i.e.
  parts of $P$ are consecutive repetitions of one of its subpath.

  \begin{proof}
    The smallest efficient tile assembly system that we found with a seed of
    size 1, has 4825 tile types, and all its terminal assemblies are of Manhattan
    radius 4845.

    To show this, we use a computer-aided proof: more specifically, we simulate
    the assembly of the tile assembly system described by the \letexp in
    Appendix \ref{partially}, yielding the assembly of Figure
    \ref{fig:partially} (also in Appendix \ref{partially}).

    Again, the full Haskell program, generating a (quite large) pdf file with
    the construction, can be found on the self-assembly wiki\footnote{
\href{http://self-assembly.net/wiki/index.php?title=Baggins-expressions}{\tt http://self-assembly.net/wiki/index.php?title=Baggins-expressions}
}.
  \end{proof}
\end{theorem}

\section{Comparisons with other models}
\label{sect:comp}

The constructions of Section \ref{sect:algos} show the intricate connections
between geometry and the computational power of temperature 1 self-assembly,
raising the question of the exact characterization of the model, from the point
of view of classical computational models. In this section, we show that we are
far from understanding these relations, and begin a broader exploration of the
influence of geometry. In Wang tilings, geometries that have been considered
previously include the hyperbolic plane~\cite{Kari-2007,Margenstern-2007} and
Cayley graphs of Baumslag-Solitar groups~\cite{Aubrun-2013,ballier-2013}.

From the self-assembly side, the models and underlying graphs that we considered are
the following:

\begin{itemize}
\item Temperature 1 tile assembly, on $\mathbb{Z}^2$.
\item Temperature 1 tile assembly, on the Cayley graph of Baumslag-Solitar
  groups.
\item Temperature 1 tile assembly, on the hyperbolic plane.
\end{itemize}

From the ``classical'' side, the computational models that we considered are the
following:

\begin{itemize}
\item Finite automata
\item Regular tree automata
\item Pushdown automata
\item Turing machines
\end{itemize}

The results shown on Figure \ref{fig:comp} are proven in Appendix \ref{app:comp}.

\begin{figure}[ht]
\begin{center}
\includegraphics[scale=0.6]{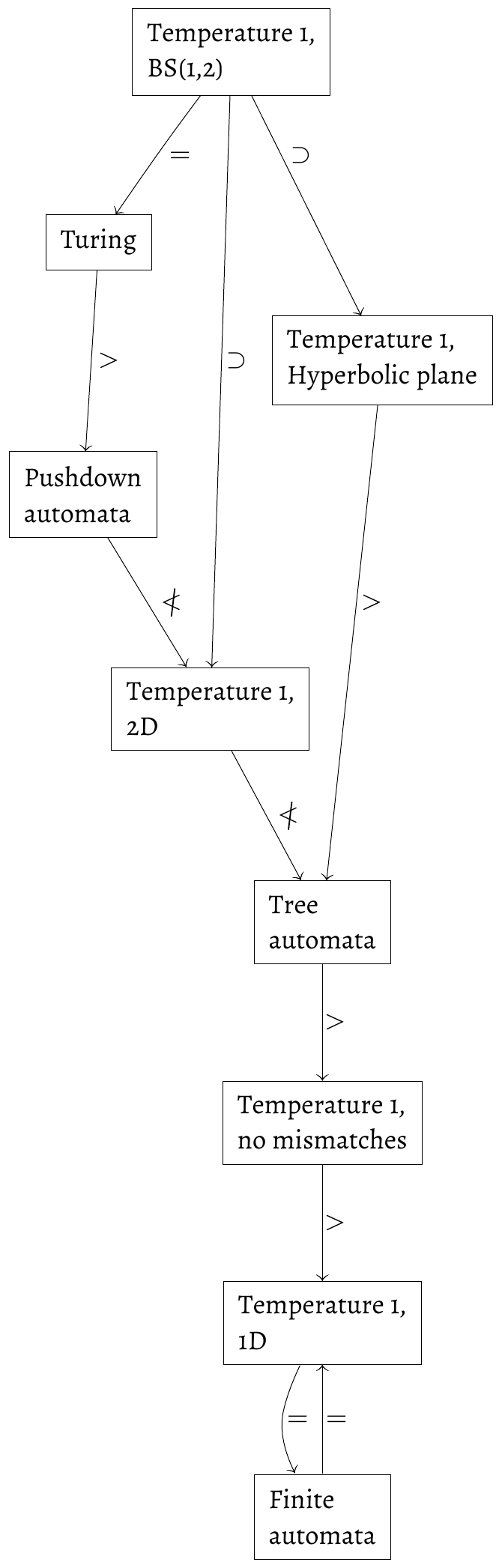}
\end{center}
\caption{Summary of the comparisons of Section \ref{app:comp}. On this graph,
  an arrow from $A$ to $B$, labeled with relation $\mathcal{R}$ means $A\mathcal{R}B$.}
\label{fig:comp}
\end{figure}

\section{Open problems and discussion}

Despite our efficient constructions, planar temperature 1 tile assembly model
does not seem capable of Turing computation. Finding the limits of these
constructions would give us a greater understanding of these processes,
ubiquitous in natural systems:

\begin{openproblem}
  What is the largest integer $s$, such that all the terminal assemblies of a
  tile assembly system with $n$ tiles and a single-tile seed, are of size $s$?
\end{openproblem}

Another question, left open by Section \ref{sect:comp}, is the exact
characterization of this model, in terms of classical models.

\newpage
\appendix
\section{A first efficient algorithm}
\label{firsteff}
\input{efficient.tex}

\section{A more general scheme}
\label{general}
\input{general.tex}

\section{A partially pumpable path}
\label{partially}

This program is slightly more complex than those of Sections \ref{firsteff} and
\ref{general}. We tried to stick to basic parts of Haskell syntax; the main
things that need to be explained are the following:
\begin{itemize}
\item the ``let'' syntax we use here is the Haskell way of defining variables, and
  is not related to the $\e{let}$ construct of \letexp{}s.
\item for reasons of efficiency, we need a new instruction called {\tt
    discreteVect}.  It is built using {\tt movex} and {\tt movey} instructions,
  combined in an efficient way.
\item {\tt quot} means ``quotient''.
\end{itemize}

\input{partially.tex}

\begin{figure}[!ht]
\begin{center}
\includegraphics[scale=0.5]{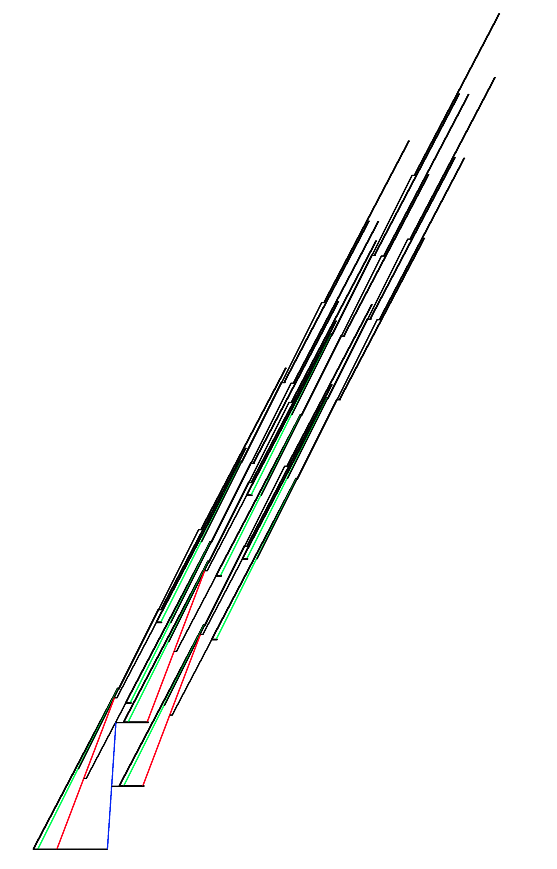}
\end{center}
\caption{A partially pumpable efficient path. The three successive partially
  pumped parts are colored in blue, red and green, successively. This image is
  rasterized for size reasons, please run the program above for a vector
  version.}
\label{fig:partially}
\end{figure}

\section{A printable version of Figure \ref{figure:efficient}}
\label{printable-eff3}
\begin{center}
\scalebox{4}{
\begin{tikzpicture}
\input{eff3}
\end{tikzpicture}
}
\end{center}

\section{Comparison with other models}
\label{app:comp}

In order to compare various settings of non-cooperative self-assembly with
classical machines from automata theory, we first introduce a notion of
language for a tile assembly system:

\begin{definition}
  \label{def:treetile}
  Let $\mathcal{T}=(T,\sigma,1)$ be a temperature 1 tile assembly system where
  $\sigma$ is single-tile seed assembly. We call $\mathcal{L}(\mathcal{T})$, the
  \emph{language} of $\mathcal{T}$, the tree language recognized by the following
  tree grammar:

  \begin{itemize}
  \item For each tile $t\in T$, with glues $t_\No$ on the north,
    $t_\Ea$ on the east, $t_\So$ on the south, and $t_\We$ on the west,
    $\mathcal{A}$ has the four following nonterminals:

    \begin{eqnarray*}
      N_{t_\No}&\rightarrow&\No(E_{t_\Ea},S_{t_\So},W_{t_\We})\\
      E_{t_\Ea}&\rightarrow&\Ea(S_{t_\So},W_{t_\We},N_{t_\No})\\
      S_{t_\So}&\rightarrow&\So(W_{t_\We},N_{t_\No},E_{t_\Ea})\\
      W_{t_\We}&\rightarrow&\We(N_{t_\No},E_{t_\Ea},S_{t_\So})
    \end{eqnarray*}

  \item Moreover, for each glue $g$ appearing on the north (\resp south, west
    and east side) of some tile of $T$, add a terminal symbol $n_g$ (\resp
    $s_g$, $w_g$, $e_g$) to the grammar, and the following rules:
    \begin{eqnarray*}
      N_{g}&\rightarrow&n_g\\
      E_{g}&\rightarrow&e_g\\
      S_{g}&\rightarrow&s_g\\
      W_{g}&\rightarrow&w_g
    \end{eqnarray*}

  \item Finally, add a nonterminal symbol $S$, and the following rule:
    \begin{eqnarray*}
      S&\rightarrow&\Sigma(N_{\sigma_\No},E_{\sigma_\Ea},S_{\sigma_\So},W_{\sigma_\We})
    \end{eqnarray*}
    Where $\sigma_\No$, $\sigma_\Ea$, $\sigma_\So$ and $\sigma_\We$ are the
    north, east, south and west glues of the unique tile of $\sigma$, respectively.
  \end{itemize}

\end{definition}

\begin{definition}

  Let $\mathcal{T}=(T,\sigma,1)$ be a temperature 1 tile assembly system.
  A term $t$ of $\mathcal{L}(\mathcal{T})$ \emph{describes} the following
  assembly sequence:

  \begin{itemize}
    \item From $\Sigma(N_{\sigma_\No},E_{\sigma_\Ea},S_{\sigma_\So},W_{\sigma_\We})$,
      concatenate the four assembly sequences obtained from
      $N_{\sigma_\No}$,
      $E_{\sigma_\Ea}$, $S_{\sigma_\So}$, $W_{\sigma_\We}$, successively.

    \item Let $\alpha(x,y,n,\No(E_{t_\Ea},S_{t_\So},W_{t_\We}))$ be concatenation
      of the following sequences:

      \begin{itemize}
        \item the assembly of the unique tile type $t\in T$ with north glue $n$,
          east glue $t_\Ea$, south glue $t_\So$ and west glue $t_\We$, at position
          $(x,y)$.
        \item assembly sequence $\alpha(x+1,y,t_\Ea,E_{t_\Ea})$.
        \item assembly sequence $\alpha(x-1,y,t_\We,W_{t_\We})$.
        \item assembly sequence $\alpha(x,y-1,t_\So,S_{t_\So})$.
      \end{itemize}

    \item Similarly for
      $\alpha(x,y,e,\Ea(S_{t_\So},W_{t_\We},N_{t_\No}))$,
      $\alpha(x,y,s,\So(W_{t_\We},N_{t_\No},E_{t_\Ea}))$,
      and $\alpha(x,y,w,\We(N_{t_\No},E_{t_\Ea},S_{t_\So}))$.

    \item For terminals $t$ of the form $n_g,s_g,e_g$ or $w_g$, let
      $\alpha(x,y,g,t)$ be the empty assembly sequence.
    \end{itemize}

    By extension, if this assembly sequence results in a producible assembly
    $a\in\prodasm{\mathcal{T}}$, we say that $t$ \emph{describes} $a$. Moreover,
    if all the terms of some tree language $L$ describe a producible assembly of
    $\mathcal{T}$, and all producible assemblies of $\mathcal{T}$ are described
    by some term $t\in L$, we say that \emph{$L$ describes
      $\prodasm{\mathcal{T}}$}.

    When all the nodes of terms of $\mathcal{L}(\mathcal{T})$ have at most one
    nonterminal child, this tree language is also a word language, over alphabet
    $T$.

\end{definition}

\begin{proposition}

  Let $A$ be a non-deterministic finite automaton on alphabet $S$. There is a
  (one-dimensional) tile assembly system $\mathcal{T}_A=(T_A,\sigma_A,1)$ such
  that $\mathcal{L}(A)$ describes $\termasm{\mathcal{T}_A}$.

  \begin{proof}
    Let $A=(Q,\Sigma,\Delta,q_0,F)$ be any non-deterministic finite automaton,
    with $Q$ its set of states, $\Sigma$ its alphabet,
    $\Delta \in Q\times\Sigma\times Q$ its transition relation,
    $q_0$ its start state and $F$ its set of final states.

    We build an ``equivalent'' temperature 1 tile assembly system
    $\mathcal{T}_A=(T_A,\sigma_A,1)$, where
    $T_A$ is a tileset with glue colors from $Q$, by letting:

    \begin{itemize}
    \item $t_\sigma$ be a tile with exactly one non-zero strength glue, on its
      east side, with color $q_0$.
    \item for each $(q,s,q')\in\Delta$, $\delta_(q,s,q')$ be a tile with color
      $q$ on its west side, $q'$ on its east side, and $s$ on its north side.
    \item for each $q\in F$, $f_q$ be a tile with color $q$ on its east side,
      and no other non-zero strength glue.
    \end{itemize}

    Then, let $T_a=\{t_\sigma\}\cup\{\delta_{(q,s,q')} |
    (q,s,q')\in\Delta\}\cup\{f_q | q\in F\}$, and $\sigma_A$ be an assembly with
    exactly one tile of type $t_\sigma$, at position $(0,0)$.

    Clearly, the language $\mathcal{L}(A)$ recognized by $A$ describes the
    terminal assemblies of $\mathcal{T}_A=(T_A,\sigma_A,1)$.
  \end{proof}

\end{proposition}

\begin{proposition}

  For any temperature 1 tile assembly system $\mathcal{T}=(T,\sigma,1)$ without
  mismatches, and such that $\sigma$ is a connected assembly, there is a
  nondeterministic top-down tree automaton whose language describes
  $\prodasm{\mathcal{T}}$.

  \begin{proof}

    Clearly, since there are no mismatches in the productions of $\mathcal{T}$,
    every assembly described by $\mathcal{L}(\mathcal{T})$ is producible by
    $\mathcal{T}$. The other direction (producible assemblies of $\mathcal{T}$
    are described by $\mathcal{L}(\mathcal{T})$ is immediate.

  \end{proof}

\end{proposition}

\begin{proposition}
  There is a temperature 1 tile assembly system $\mathcal{T}$ such that
  $\mathcal{L}(\mathcal{T})$ describes assembly sequences not producible
  by $\mathcal{T}$.

  \begin{proof}
    Let $T$ be the following tileset:
    $$T=\left\{
    t_0=\scalebox{0.7}{
      \begin{tikzpicture}[baseline={(0,0.2)}]
        \draw(0,0)rectangle(1,1);
        \draw(0.5,1)--(0.5,0.9)node[anchor=north,inner sep=1pt]{$a$};
      \end{tikzpicture}
    },
    t_1=\scalebox{0.7}{
      \begin{tikzpicture}[baseline={(0,0.2)}]
      \draw(0,0)rectangle(1,1);
      \draw(0.5,1)--(0.5,0.9)node[anchor=north,inner sep=1pt]{$a$};
      \draw(0.5,0)--(0.5,0.1)node[anchor=south,inner sep=1pt]{$a$};
      \end{tikzpicture}
    },
    t_2=\scalebox{0.7}{
      \begin{tikzpicture}[baseline={(0,0.2)}]
      \draw(0,0)rectangle(1,1);
      \draw(0.5,0)--(0.5,0.1)node[anchor=south,inner sep=1pt]{$a$};
      \draw(1,0.5)--(0.9,0.5)node[anchor=east,inner sep=1pt]{$a$};
      \end{tikzpicture}
    },
    t_3=\scalebox{0.7}{
      \begin{tikzpicture}[baseline={(0,0.2)}]
      \draw(0,0)rectangle(1,1);
      \draw(1,0.5)--(0.9,0.5)node[anchor=east,inner sep=1pt]{$a$};
      \draw(0,0.5)--(0.1,0.5)node[anchor=west,inner sep=1pt]{$a$};
      \end{tikzpicture}
    },
    t_4=\scalebox{0.7}{
      \begin{tikzpicture}[baseline={(0,0.2)}]
      \draw(0,0)rectangle(1,1);
      \draw(0.5,0)--(0.5,0.1)node[anchor=south,inner sep=1pt]{$b$};
      \draw(0,0.5)--(0.1,0.5)node[anchor=west,inner sep=1pt]{$a$};
      \end{tikzpicture}
    },
    t_5=\scalebox{0.7}{
      \begin{tikzpicture}[baseline={(0,0.2)}]
      \draw(0,0)rectangle(1,1);
      \draw(0.5,1)--(0.5,0.9)node[anchor=north,inner sep=1pt]{$b$};
      \draw(0.5,0)--(0.5,0.1)node[anchor=south,inner sep=1pt]{$b$};
      \end{tikzpicture}
    },
    t_6=\scalebox{0.7}{
      \begin{tikzpicture}[baseline={(0,0.2)}]
      \draw(0,0)rectangle(1,1);
      \draw(0.5,1)--(0.5,0.9)node[anchor=north,inner sep=1pt]{$b$};
      \draw(0,0.5)--(0.1,0.5)node[anchor=west,inner sep=1pt]{$c$};
      \end{tikzpicture}
    },
    t_7=\scalebox{0.7}{
      \begin{tikzpicture}[baseline={(0,0.2)}]
      \draw(0,0)rectangle(1,1);
      \draw(1,0.5)--(0.9,0.5)node[anchor=east,inner sep=1pt]{$c$};
      \draw(0,0.5)--(0.1,0.5)node[anchor=west,inner sep=1pt]{$c$};
      \end{tikzpicture}
    }\right\}$$

  Let $\sigma$ be the assembly with a single tile of type $t_0$.

  We claim that for $\mathcal{T}=(T,\sigma,1)$, $\mathcal{L}(\mathcal{T})$
  describes assembly sequences not representing any assembly. First, since all
  the tiles of $T$ can attach to at most two tiles, we can completely describe
  assembly sequences as words on $T$. Let $L$ be the language of all assembly sequences
  ($L$ is therefore a word language on alphabet $T$).

  Since $\mathcal{L}(\mathcal{T})$ is a regular tree language, $L$ is a regular
  language, and is therefore recognized by a deterministic finite automaton $A$.
  Let $n$ be the number of states of $A$, and let $u=t_0t_1^{n}t_2t_4t_5^{n+1}t_6t_7^{10}$.
  Moreover, for $i\in\{0,1,\ldots,|u|-1\}$, let $a_i$ be the state in which $A$ is
  just before letter $u_i$. Since there are $n+1$ occurrences of $t_5$ in $u$, at least
  two distinct indices $i$ and $j$, in subword $t_5^{n+1}$ of $u$, are such that $a_i=a_j$.

  This means that the following word, which does not described any production
  of $\mathcal{T}$, is recognized: $t_0t_1^{n}t_2t_4t_5^{n+1-b+a}t_6t_7^{10}$.

  \end{proof}
\end{proposition}

\begin{proposition}

  There is a temperature 1 tile assembly system $\mathcal{T}=(T,\sigma,1)$ such
  that $\mathcal{L}(\mathcal{T})$ is a non-context-free word language on
  alphabet $T$.

  \begin{proof}
    Let $T$ be the following tileset:
    $$\begin{array}{cc}
      T=&\left\{
        t_0=\scalebox{0.7}{
          \begin{tikzpicture}[baseline={(0,0.2)}]
            \draw(0,0)rectangle(1,1);
            \draw(0.5,1)--(0.5,0.9)node[anchor=north,inner sep=1pt]{$a_0$};
          \end{tikzpicture}
        },
        t_1=\scalebox{0.7}{
          \begin{tikzpicture}[baseline={(0,0.2)}]
            \draw(0,0)rectangle(1,1);
            \draw(0.5,1)--(0.5,0.9)node[anchor=north,inner sep=1pt]{$a_1$};
            \draw(0.5,0)--(0.5,0.1)node[anchor=south,inner sep=1pt]{$a_0$};
          \end{tikzpicture}
        },
        t_2=\scalebox{0.7}{
          \begin{tikzpicture}[baseline={(0,0.2)}]
            \draw(0,0)rectangle(1,1);
            \draw(0.5,0)--(0.5,0.1)node[anchor=south,inner sep=1pt]{$a_1$};
            \draw(1,0.5)--(0.9,0.5)node[anchor=east,inner sep=1pt]{$b$};
          \end{tikzpicture}
        },
        t_3=\scalebox{0.7}{
          \begin{tikzpicture}[baseline={(0,0.2)}]
            \draw(0,0)rectangle(1,1);
            \draw(1,0.5)--(0.9,0.5)node[anchor=east,inner sep=1pt]{$b$};
            \draw(0,0.5)--(0.1,0.5)node[anchor=west,inner sep=1pt]{$b$};
          \end{tikzpicture}
        },
        t_4=\scalebox{0.7}{
          \begin{tikzpicture}[baseline={(0,0.2)}]
            \draw(0,0)rectangle(1,1);
            \draw(0.5,0)--(0.5,0.1)node[anchor=south,inner sep=1pt]{$c_2$};
            \draw(0,0.5)--(0.1,0.5)node[anchor=west,inner sep=1pt]{$b$};
          \end{tikzpicture}
        },
        t_5=\scalebox{0.7}{
          \begin{tikzpicture}[baseline={(0,0.2)}]
            \draw(0,0)rectangle(1,1);
            \draw(0.5,1)--(0.5,0.9)node[anchor=north,inner sep=1pt]{$c_2$};
            \draw(0.5,0)--(0.5,0.1)node[anchor=south,inner sep=1pt]{$c_1$};
          \end{tikzpicture}
        },\right.\\&\left.
        t_6=\scalebox{0.7}{
          \begin{tikzpicture}[baseline={(0,0.2)}]
            \draw(0,0)rectangle(1,1);
            \draw(0.5,1)--(0.5,0.9)node[anchor=north,inner sep=1pt]{$c_1$};
            \draw(0,0.5)--(0.1,0.5)node[anchor=west,inner sep=1pt]{$d$};
          \end{tikzpicture}
        },
        t_7=\scalebox{0.7}{
      \begin{tikzpicture}[baseline={(0,0.2)}]
      \draw(0,0)rectangle(1,1);
      \draw(1,0.5)--(0.9,0.5)node[anchor=east,inner sep=1pt]{$d$};
      \draw(0,0.5)--(0.1,0.5)node[anchor=west,inner sep=1pt]{$d$};
      \end{tikzpicture}
    },
    t_8=\scalebox{0.7}{
      \begin{tikzpicture}[baseline={(0,0.2)}]
      \draw(0,0)rectangle(1,1);
      \draw(0.5,1)--(0.5,0.9)node[anchor=north,inner sep=1pt]{$e$};
      \draw(1,0.5)--(0.9,0.5)node[anchor=east,inner sep=1pt]{$d$};
      \end{tikzpicture}
    },
    t_{9}=\scalebox{0.7}{
      \begin{tikzpicture}[baseline={(0,0.2)}]
      \draw(0,0)rectangle(1,1);
      \draw(0.5,0)--(0.5,0.1)node[anchor=south,inner sep=1pt]{$e$};
      \draw(1,0.5)--(0.9,0.5)node[anchor=east,inner sep=1pt]{$f$};
      \end{tikzpicture}
    },
    t_{10}=\scalebox{0.7}{
      \begin{tikzpicture}[baseline={(0,0.2)}]
      \draw(0,0)rectangle(1,1);
      \draw(1,0.5)--(0.9,0.5)node[anchor=east,inner sep=1pt]{$f$};
      \draw(0,0.5)--(0.1,0.5)node[anchor=west,inner sep=1pt]{$f$};
      \end{tikzpicture}
    }\right\}
  \end{array}$$

  Since all tiles of $T$ have exactly two sides of non-zero strength, the tree
  language $\mathcal{L}(\mathcal{T})$ is actually also a word language, on
  alphabet $T$. However, the language $L$ of the productions of $T$ is the union
  of the language $M$ describing the terminal assemblies of $\mathcal{T}$, with
  all the prefixes of these assemblies. Formally, $M$ is the following language:

  $$M=\{t_0t_1t_2t_3^at_4t_5t_6t_7^bt_8t_9t_{10}^c | a>b\geq c\}
  \cup \{t_0t_1t_2t_3^at_4t_5t_6t_7^a | a\in\mathbb{N}\}$$

  Moreover, by the pumping Lemma on pushdown automata, this means if $L$ were
  context-free, then it would also contain words of the form
  $t_0t_1t_2t_3^at_4t_5t_6t_7^bt_8t_9t_{10}^c$ in which either $c>b$ or $b\geq
  a$, which is not the case. Indeed, for all $a$, $M$ contains the following word:
  $$t_0t_1t_2t_3^{a+1}t_4t_5t_6t_7^{a}t_8t_9t_{10}^{a}$$
  Therefore, the pumping lemma states that $L$ were context-free, it would also contain:
  \begin{itemize}
  \item Either $t_0t_1t_2t_3^{a+1-b}t_4t_5t_6t_7^{a-b}t_8t_9t_{10}^{a}$
    for some $b<a$. However, this word is not in $L$.
  \item Or $t_0t_1t_2t_3^{a+1}t_4t_5t_6t_7^{a+b}t_8t_9t_{10}^{a+b}$ for some $b>0$,
    which is also not in $L$.
  \item Or $t_0t_1t_2t_3^{a+1+b}t_4t_5t_6t_7^{a}t_8t_9t_{10}^{a+b}$ for some $b>0$,
    which is also not in $L$.
  \end{itemize}
  \end{proof}
\end{proposition}

\begin{definition}
  A Baumslag-Solitar group of integer parameters $m$ and $n$ is a group given by
  the following presentation (with two generators $a$ and $b$, and one relation):
  $$B(m,n)=\langle a,b\ |\ ba^m=a^nb\rangle$$
\end{definition}

\begin{proposition}

  For any Turing machine $M$ and all input $x\in\mathbb{N}$ for $M$, there is a
  tile assembly system $\mathcal{T}_{M,x}=(T_M,\sigma_{M,x},1)$ on
  Baumslag-Solitar group $B(1,2)$, and a tile $t\in T_M$, such that:
  \begin{itemize}
    \item $\sigma_{M,x}$ is recursive
    \item all terminal assemblies of $\mathcal{T}_{M,x}$ contain $t$ if
      and only if $M$ accepts $x$.
    \end{itemize}

  \begin{proof}

    This is a straightforward adaptation of the 3D construction of
    Cook, Fu and Schweller~\cite{Cook-2011}, simulating zig-zag systems (and thus Turing machines).

    The geometric intuition is that $B(1,2)$ is a ``tree of half-planes'' (see
    Figure~\ref{fig:bs}). In the construction, we will most of the time stay in
    the ``initial'' plane, i.e. the leftmost branch of the tree, and avoid
    planarity by taking another branch temporarily.

    Now, contrarily to the grid graph of $\mathbb{Z}^3$, there is no edge in
    the Cayley graph of $BS(1,2)$ between these ``half planes''.

    \begin{figure}[ht]
      \begin{center}
        \begin{tikzpicture}
          \input{bs}
        \end{tikzpicture}
      \end{center}
      \caption{Some points and relations of $BS(1,2)$. Different ``half-planes''
        are in different colors.}
      \label{fig:bs}
    \end{figure}

    However, their bit selection gadget can be adapted to $BS(1,2)$, in the way
    depicted on Figure~\ref{fig:bitsel}: the red and green paths encode a zero
    or a one. In order to read it, the orange path forks into two branches, and
    only one is allowed to pass through the encoding (the other one collides
    against a part of the encoded bit).

    \begin{figure}[ht]
      \begin{center}
        \begin{tikzpicture}
          \input{bitsel}
          \input{bitsel1}
          \begin{scope}[thick,dashed]

          \draw[orange](2,2)--(1.75,2);
          \draw[orange,->](1.75,2)--(1,2);
          \draw[orange,->](1,2)--(-1,2); %
          \begin{scope}[xshift=0.5cm]
            \draw[orange](1.25,2)..controls(1.125,2) and (1.125,2.25)..(1.25,2.25)
            ..controls(1.125,2.25) and (1.125,2.5)..(1.25,2.5)--(1,2.5);;
          \end{scope}
          \end{scope}
        \end{tikzpicture}\hspace{2cm}
        \begin{tikzpicture}
          \input{bitsel}
          \input{bitsel2}
          \begin{scope}[thick,dashed]
          \draw[orange](2,2)--(1.75,2);
          \end{scope}
          \begin{scope}[xshift=0.5cm,thick,dashed]
            \draw[orange,->](1.25,2)..controls(1.125,2) and (1.125,2.25)..(1.25,2.25)
            ..controls(1.125,2.25) and (1.125,2.5)..(1.25,2.5)--(1,2.5);;

            \draw[orange](1,2.5)..controls(0.75,2.5) and (0.75,3)..(1,3)--(0.5,3);

            \draw[orange](0.5,3)..controls(0,3) and (0,4)..(0.5,4)--(0,4);
            \draw[orange,->](0,4)--(-0.5,4);

            \draw[orange,->](-0.5,4)..controls(-1.5,4) and (-1.5,2)..(-0.5,2)--(-1.5,2);
          \end{scope}
        \end{tikzpicture}
      \end{center}
      \caption{Adapting the bit selection gadget of \cite{Cook-2011} to
        $BS(1,2)$.  In this figure, the red/green paths grow first, and encode a
        0 on the left assembly, and a 1 on the right one. The parts of the
        initial paths that are on the first ``plane'' are in red, other parts
        are in green. The orange (dashed) paths are paths from the next row,
        that read this encoding.}
      \label{fig:bitsel}
    \end{figure}

  \end{proof}

\end{proposition}

\begin{proposition}

  Let $L$ be a regular tree language of degree at most $d$. There is a
  tile assembly system $(T,\sigma,1)$ in the hyperbolic plane, where
  $|\dom\sigma|=1$, and such that $\prodasm{\mathcal{T}}$ is described by $L$.

  \begin{proof}
    The hyperbolic plane is a tree of degree $k$, along with edges between
    consecutive vertices of the same level, and an edge between the first and
    last vertices of each level (see \cite{Margenstern-2007} for more details).

    Therefore, simulating a tree automaton of degree $k$ is straightforward,
    and Definition \ref{def:treetile} allows us to conclude.
  \end{proof}
\end{proposition}

\bibliographystyle{plain}

\input{noncoop.bbl}
\end{document}

%% file: eff1.tex
\draw[use as bounding box,draw=none,fill=none](0.4000,0.0000)rectangle(1.8000,3.6000);\begin{scope}[transform canvas={scale=0.12}]\draw(11.666666666666668,0.0)rectangle(13.333333333333334,1.6666666666666667);\draw(12.5,1.6666666666666667)node[anchor=north]{1};\end{scope}

%% file: eff2.tex
\draw[use as bounding box,draw=none,fill=none](0.0000,0.0000)rectangle(1.8000,3.6000);\begin{scope}[transform canvas={scale=0.12}]\draw(0.0,20.0)rectangle(1.6666666666666667,21.666666666666668);\draw(0.8333333333333334,21.666666666666668)node[anchor=north]{20};\draw(1.6666666666666667,20.833333333333336)node[anchor=east]{19};\draw(0.0,21.666666666666668)rectangle(1.6666666666666667,23.333333333333336);\draw(0.8333333333333334,23.333333333333336)node[anchor=north]{21};\draw(0.8333333333333334,21.666666666666668)node[anchor=south]{20};\draw(0.0,23.333333333333336)rectangle(1.6666666666666667,25.0);\draw(0.8333333333333334,25.0)node[anchor=north]{22};\draw(0.8333333333333334,23.333333333333336)node[anchor=south]{21};\draw(0.0,25.0)rectangle(1.6666666666666667,26.666666666666668);\draw(0.8333333333333334,26.666666666666668)node[anchor=north]{23};\draw(0.8333333333333334,25.0)node[anchor=south]{22};\draw(1.6666666666666667,20.0)rectangle(3.3333333333333335,21.666666666666668);\draw(1.6666666666666667,20.833333333333336)node[anchor=west]{19};\draw(3.3333333333333335,20.833333333333336)node[anchor=east]{18};\draw(3.3333333333333335,20.0)rectangle(5.0,21.666666666666668);\draw(3.3333333333333335,20.833333333333336)node[anchor=west]{18};\draw(5.0,20.833333333333336)node[anchor=east]{17};\draw(5.0,20.0)rectangle(6.666666666666667,21.666666666666668);\draw(5.0,20.833333333333336)node[anchor=west]{17};\draw(6.666666666666667,20.833333333333336)node[anchor=east]{16};\draw(6.666666666666667,20.0)rectangle(8.333333333333334,21.666666666666668);\draw(6.666666666666667,20.833333333333336)node[anchor=west]{16};\draw(8.333333333333334,20.833333333333336)node[anchor=east]{15};\draw(8.333333333333334,10.0)rectangle(10.0,11.666666666666668);\draw(9.166666666666668,11.666666666666668)node[anchor=north]{9};\draw(10.0,10.833333333333334)node[anchor=east]{8};\draw(8.333333333333334,11.666666666666668)rectangle(10.0,13.333333333333334);\draw(9.166666666666668,13.333333333333334)node[anchor=north]{10};\draw(9.166666666666668,11.666666666666668)node[anchor=south]{9};\draw(8.333333333333334,13.333333333333334)rectangle(10.0,15.0);\draw(9.166666666666668,15.0)node[anchor=north]{11};\draw(9.166666666666668,13.333333333333334)node[anchor=south]{10};\draw(8.333333333333334,15.0)rectangle(10.0,16.666666666666668);\draw(9.166666666666668,16.666666666666668)node[anchor=north]{12};\draw(9.166666666666668,15.0)node[anchor=south]{11};\draw(8.333333333333334,16.666666666666668)rectangle(10.0,18.333333333333336);\draw(9.166666666666668,18.333333333333336)node[anchor=north]{13};\draw(9.166666666666668,16.666666666666668)node[anchor=south]{12};\draw(8.333333333333334,18.333333333333336)rectangle(10.0,20.0);\draw(9.166666666666668,20.0)node[anchor=north]{14};\draw(9.166666666666668,18.333333333333336)node[anchor=south]{13};\draw(8.333333333333334,20.0)rectangle(10.0,21.666666666666668);\draw(8.333333333333334,20.833333333333336)node[anchor=west]{15};\draw(9.166666666666668,21.666666666666668)node[anchor=north]{36};\draw(9.166666666666668,20.0)node[anchor=south]{14};\draw(10.0,10.0)rectangle(11.666666666666668,11.666666666666668);\draw(10.0,10.833333333333334)node[anchor=west]{8};\draw(11.666666666666668,10.833333333333334)node[anchor=east]{7};\draw(11.666666666666668,0.0)rectangle(13.333333333333334,1.6666666666666667);\draw(12.5,1.6666666666666667)node[anchor=north]{1};\draw(11.666666666666668,1.6666666666666667)rectangle(13.333333333333334,3.3333333333333335);\draw(12.5,3.3333333333333335)node[anchor=north]{2};\draw(12.5,1.6666666666666667)node[anchor=south]{1};\draw(11.666666666666668,3.3333333333333335)rectangle(13.333333333333334,5.0);\draw(12.5,5.0)node[anchor=north]{3};\draw(12.5,3.3333333333333335)node[anchor=south]{2};\draw(11.666666666666668,5.0)rectangle(13.333333333333334,6.666666666666667);\draw(12.5,6.666666666666667)node[anchor=north]{4};\draw(12.5,5.0)node[anchor=south]{3};\draw(11.666666666666668,6.666666666666667)rectangle(13.333333333333334,8.333333333333334);\draw(12.5,8.333333333333334)node[anchor=north]{5};\draw(12.5,6.666666666666667)node[anchor=south]{4};\draw(11.666666666666668,8.333333333333334)rectangle(13.333333333333334,10.0);\draw(12.5,10.0)node[anchor=north]{6};\draw(12.5,8.333333333333334)node[anchor=south]{5};\draw(11.666666666666668,10.0)rectangle(13.333333333333334,11.666666666666668);\draw(11.666666666666668,10.833333333333334)node[anchor=west]{7};\draw(13.333333333333334,10.833333333333334)node[anchor=east]{32};\draw(12.5,10.0)node[anchor=south]{6};\end{scope}

%% file: eff0.tex
\draw[use as bounding box,draw=none,fill=none](0.0000,0.0000)rectangle(1.8000,5.4000);\begin{scope}[transform canvas={scale=0.12}]\draw(0.0,20.0)rectangle(1.6666666666666667,21.666666666666668);\draw(0.8333333333333334,21.666666666666668)node[anchor=north]{20};\draw(1.6666666666666667,20.833333333333336)node[anchor=east]{19};\draw(0.0,21.666666666666668)rectangle(1.6666666666666667,23.333333333333336);\draw(0.8333333333333334,23.333333333333336)node[anchor=north]{21};\draw(0.8333333333333334,21.666666666666668)node[anchor=south]{20};\draw(0.0,23.333333333333336)rectangle(1.6666666666666667,25.0);\draw(0.8333333333333334,25.0)node[anchor=north]{22};\draw(0.8333333333333334,23.333333333333336)node[anchor=south]{21};\draw(0.0,25.0)rectangle(1.6666666666666667,26.666666666666668);\draw(0.8333333333333334,26.666666666666668)node[anchor=north]{23};\draw(0.8333333333333334,25.0)node[anchor=south]{22};\draw(0.0,26.666666666666668)rectangle(1.6666666666666667,28.333333333333336);\draw(0.8333333333333334,28.333333333333336)node[anchor=north]{24};\draw(0.8333333333333334,26.666666666666668)node[anchor=south]{23};\draw(0.0,28.333333333333336)rectangle(1.6666666666666667,30.0);\draw(1.6666666666666667,29.166666666666668)node[anchor=east]{25};\draw(0.8333333333333334,28.333333333333336)node[anchor=south]{24};\draw(1.6666666666666667,20.0)rectangle(3.3333333333333335,21.666666666666668);\draw(1.6666666666666667,20.833333333333336)node[anchor=west]{19};\draw(3.3333333333333335,20.833333333333336)node[anchor=east]{18};\draw(1.6666666666666667,21.666666666666668)rectangle(3.3333333333333335,23.333333333333336);\draw(2.5,23.333333333333336)node[anchor=north]{29};\draw(3.3333333333333335,22.5)node[anchor=east]{30};\draw(1.6666666666666667,23.333333333333336)rectangle(3.3333333333333335,25.0);\draw(2.5,25.0)node[anchor=north]{28};\draw(3.3333333333333335,24.166666666666668)node[anchor=east]{35};\draw(2.5,23.333333333333336)node[anchor=south]{29};\draw(1.6666666666666667,25.0)rectangle(3.3333333333333335,26.666666666666668);\draw(2.5,26.666666666666668)node[anchor=north]{27};\draw(2.5,25.0)node[anchor=south]{28};\draw(1.6666666666666667,26.666666666666668)rectangle(3.3333333333333335,28.333333333333336);\draw(2.5,28.333333333333336)node[anchor=north]{26};\draw(2.5,26.666666666666668)node[anchor=south]{27};\draw(1.6666666666666667,28.333333333333336)rectangle(3.3333333333333335,30.0);\draw(1.6666666666666667,29.166666666666668)node[anchor=west]{25};\draw(2.5,28.333333333333336)node[anchor=south]{26};\draw(3.3333333333333335,20.0)rectangle(5.0,21.666666666666668);\draw(3.3333333333333335,20.833333333333336)node[anchor=west]{18};\draw(5.0,20.833333333333336)node[anchor=east]{17};\draw(3.3333333333333335,21.666666666666668)rectangle(5.0,23.333333333333336);\draw(3.3333333333333335,22.5)node[anchor=west]{30};\draw(5.0,22.5)node[anchor=east]{31};\draw(5.0,20.0)rectangle(6.666666666666667,21.666666666666668);\draw(5.0,20.833333333333336)node[anchor=west]{17};\draw(6.666666666666667,20.833333333333336)node[anchor=east]{16};\draw(5.0,21.666666666666668)rectangle(6.666666666666667,23.333333333333336);\draw(5.0,22.5)node[anchor=west]{31};\draw(5.833333333333334,23.333333333333336)node[anchor=north]{3};\draw(5.0,23.333333333333336)rectangle(6.666666666666667,25.0);\draw(5.833333333333334,25.0)node[anchor=north]{4};\draw(5.833333333333334,23.333333333333336)node[anchor=south]{3};\draw(5.0,25.0)rectangle(6.666666666666667,26.666666666666668);\draw(5.833333333333334,26.666666666666668)node[anchor=north]{5};\draw(5.833333333333334,25.0)node[anchor=south]{4};\draw(5.0,26.666666666666668)rectangle(6.666666666666667,28.333333333333336);\draw(5.833333333333334,28.333333333333336)node[anchor=north]{6};\draw(5.833333333333334,26.666666666666668)node[anchor=south]{5};\draw(5.0,28.333333333333336)rectangle(6.666666666666667,30.0);\draw(5.0,29.166666666666668)node[anchor=west]{7};\draw(6.666666666666667,29.166666666666668)node[anchor=east]{32};\draw(5.833333333333334,28.333333333333336)node[anchor=south]{6};\draw(5.0,30.0)rectangle(6.666666666666667,31.666666666666668);\draw(5.833333333333334,31.666666666666668)node[anchor=north]{21};\draw(6.666666666666667,30.833333333333336)node[anchor=east]{34};\draw(5.0,31.666666666666668)rectangle(6.666666666666667,33.333333333333336);\draw(5.833333333333334,33.333333333333336)node[anchor=north]{22};\draw(5.833333333333334,31.666666666666668)node[anchor=south]{21};\draw(5.0,33.333333333333336)rectangle(6.666666666666667,35.0);\draw(5.833333333333334,35.0)node[anchor=north]{23};\draw(5.833333333333334,33.333333333333336)node[anchor=south]{22};\draw(5.0,35.0)rectangle(6.666666666666667,36.66666666666667);\draw(5.833333333333334,36.66666666666667)node[anchor=north]{24};\draw(5.833333333333334,35.0)node[anchor=south]{23};\draw(5.0,36.66666666666667)rectangle(6.666666666666667,38.333333333333336);\draw(6.666666666666667,37.5)node[anchor=east]{25};\draw(5.833333333333334,36.66666666666667)node[anchor=south]{24};\draw(6.666666666666667,20.0)rectangle(8.333333333333334,21.666666666666668);\draw(6.666666666666667,20.833333333333336)node[anchor=west]{16};\draw(8.333333333333334,20.833333333333336)node[anchor=east]{15};\draw(6.666666666666667,28.333333333333336)rectangle(8.333333333333334,30.0);\draw(6.666666666666667,29.166666666666668)node[anchor=west]{32};\draw(7.5,30.0)node[anchor=north]{33};\draw(6.666666666666667,30.0)rectangle(8.333333333333334,31.666666666666668);\draw(6.666666666666667,30.833333333333336)node[anchor=west]{34};\draw(7.5,30.0)node[anchor=south]{33};\draw(6.666666666666667,31.666666666666668)rectangle(8.333333333333334,33.333333333333336);\draw(7.5,33.333333333333336)node[anchor=north]{28};\draw(8.333333333333334,32.5)node[anchor=east]{35};\draw(7.5,31.666666666666668)node[anchor=south]{29};\draw(6.666666666666667,33.333333333333336)rectangle(8.333333333333334,35.0);\draw(7.5,35.0)node[anchor=north]{27};\draw(7.5,33.333333333333336)node[anchor=south]{28};\draw(6.666666666666667,35.0)rectangle(8.333333333333334,36.66666666666667);\draw(7.5,36.66666666666667)node[anchor=north]{26};\draw(7.5,35.0)node[anchor=south]{27};\draw(6.666666666666667,36.66666666666667)rectangle(8.333333333333334,38.333333333333336);\draw(6.666666666666667,37.5)node[anchor=west]{25};\draw(7.5,36.66666666666667)node[anchor=south]{26};\draw(6.666666666666667,38.333333333333336)rectangle(8.333333333333334,40.0);\draw(7.5,40.0)node[anchor=north]{22};\draw(8.333333333333334,39.16666666666667)node[anchor=east]{37};\draw(6.666666666666667,40.0)rectangle(8.333333333333334,41.66666666666667);\draw(7.5,41.66666666666667)node[anchor=north]{23};\draw(7.5,40.0)node[anchor=south]{22};\draw(6.666666666666667,41.66666666666667)rectangle(8.333333333333334,43.333333333333336);\draw(7.5,43.333333333333336)node[anchor=north]{24};\draw(7.5,41.66666666666667)node[anchor=south]{23};\draw(6.666666666666667,43.333333333333336)rectangle(8.333333333333334,45.0);\draw(8.333333333333334,44.16666666666667)node[anchor=east]{25};\draw(7.5,43.333333333333336)node[anchor=south]{24};\draw(8.333333333333334,10.0)rectangle(10.0,11.666666666666668);\draw(9.166666666666668,11.666666666666668)node[anchor=north]{9};\draw(10.0,10.833333333333334)node[anchor=east]{8};\draw(8.333333333333334,11.666666666666668)rectangle(10.0,13.333333333333334);\draw(9.166666666666668,13.333333333333334)node[anchor=north]{10};\draw(9.166666666666668,11.666666666666668)node[anchor=south]{9};\draw(8.333333333333334,13.333333333333334)rectangle(10.0,15.0);\draw(9.166666666666668,15.0)node[anchor=north]{11};\draw(9.166666666666668,13.333333333333334)node[anchor=south]{10};\draw(8.333333333333334,15.0)rectangle(10.0,16.666666666666668);\draw(9.166666666666668,16.666666666666668)node[anchor=north]{12};\draw(9.166666666666668,15.0)node[anchor=south]{11};\draw(8.333333333333334,16.666666666666668)rectangle(10.0,18.333333333333336);\draw(9.166666666666668,18.333333333333336)node[anchor=north]{13};\draw(9.166666666666668,16.666666666666668)node[anchor=south]{12};\draw(8.333333333333334,18.333333333333336)rectangle(10.0,20.0);\draw(9.166666666666668,20.0)node[anchor=north]{14};\draw(9.166666666666668,18.333333333333336)node[anchor=south]{13};\draw(8.333333333333334,20.0)rectangle(10.0,21.666666666666668);\draw(8.333333333333334,20.833333333333336)node[anchor=west]{15};\draw(9.166666666666668,21.666666666666668)node[anchor=north]{36};\draw(9.166666666666668,20.0)node[anchor=south]{14};\draw(8.333333333333334,31.666666666666668)rectangle(10.0,33.333333333333336);\draw(8.333333333333334,32.5)node[anchor=west]{35};\draw(9.166666666666668,33.333333333333336)node[anchor=north]{12};\draw(8.333333333333334,33.333333333333336)rectangle(10.0,35.0);\draw(9.166666666666668,35.0)node[anchor=north]{13};\draw(9.166666666666668,33.333333333333336)node[anchor=south]{12};\draw(8.333333333333334,35.0)rectangle(10.0,36.66666666666667);\draw(9.166666666666668,36.66666666666667)node[anchor=north]{14};\draw(9.166666666666668,35.0)node[anchor=south]{13};\draw(8.333333333333334,36.66666666666667)rectangle(10.0,38.333333333333336);\draw(8.333333333333334,37.5)node[anchor=west]{15};\draw(9.166666666666668,38.333333333333336)node[anchor=north]{36};\draw(9.166666666666668,36.66666666666667)node[anchor=south]{14};\draw(8.333333333333334,38.333333333333336)rectangle(10.0,40.0);\draw(8.333333333333334,39.16666666666667)node[anchor=west]{37};\draw(9.166666666666668,38.333333333333336)node[anchor=south]{36};\draw(8.333333333333334,40.0)rectangle(10.0,41.66666666666667);\draw(9.166666666666668,41.66666666666667)node[anchor=north]{27};\draw(9.166666666666668,40.0)node[anchor=south]{28};\draw(8.333333333333334,41.66666666666667)rectangle(10.0,43.333333333333336);\draw(9.166666666666668,43.333333333333336)node[anchor=north]{26};\draw(9.166666666666668,41.66666666666667)node[anchor=south]{27};\draw(8.333333333333334,43.333333333333336)rectangle(10.0,45.0);\draw(8.333333333333334,44.16666666666667)node[anchor=west]{25};\draw(9.166666666666668,43.333333333333336)node[anchor=south]{26};\draw(10.0,10.0)rectangle(11.666666666666668,11.666666666666668);\draw(10.0,10.833333333333334)node[anchor=west]{8};\draw(11.666666666666668,10.833333333333334)node[anchor=east]{7};\draw(11.666666666666668,0.0)rectangle(13.333333333333334,1.6666666666666667);\draw(12.5,1.6666666666666667)node[anchor=north]{1};\draw(11.666666666666668,1.6666666666666667)rectangle(13.333333333333334,3.3333333333333335);\draw(12.5,3.3333333333333335)node[anchor=north]{2};\draw(12.5,1.6666666666666667)node[anchor=south]{1};\draw(11.666666666666668,3.3333333333333335)rectangle(13.333333333333334,5.0);\draw(12.5,5.0)node[anchor=north]{3};\draw(12.5,3.3333333333333335)node[anchor=south]{2};\draw(11.666666666666668,5.0)rectangle(13.333333333333334,6.666666666666667);\draw(12.5,6.666666666666667)node[anchor=north]{4};\draw(12.5,5.0)node[anchor=south]{3};\draw(11.666666666666668,6.666666666666667)rectangle(13.333333333333334,8.333333333333334);\draw(12.5,8.333333333333334)node[anchor=north]{5};\draw(12.5,6.666666666666667)node[anchor=south]{4};\draw(11.666666666666668,8.333333333333334)rectangle(13.333333333333334,10.0);\draw(12.5,10.0)node[anchor=north]{6};\draw(12.5,8.333333333333334)node[anchor=south]{5};\draw(11.666666666666668,10.0)rectangle(13.333333333333334,11.666666666666668);\draw(11.666666666666668,10.833333333333334)node[anchor=west]{7};\draw(13.333333333333334,10.833333333333334)node[anchor=east]{32};\draw(12.5,10.0)node[anchor=south]{6};\end{scope}

%% file: eff4.tex
\draw[use as bounding box,draw=none,fill=none](0.0000,0.0000)rectangle(2.0000,12.4000);\begin{scope}[transform canvas={scale=0.12}]\draw(0.0,43.333333333333336)rectangle(1.6666666666666667,45.0);\draw(0.8333333333333334,45.0)node[anchor=north]{34};\draw(1.6666666666666667,44.16666666666667)node[anchor=east]{33};\draw(0.0,45.0)rectangle(1.6666666666666667,46.66666666666667);\draw(0.8333333333333334,46.66666666666667)node[anchor=north]{35};\draw(0.8333333333333334,45.0)node[anchor=south]{34};\draw(0.0,46.66666666666667)rectangle(1.6666666666666667,48.333333333333336);\draw(0.8333333333333334,48.333333333333336)node[anchor=north]{36};\draw(0.8333333333333334,46.66666666666667)node[anchor=south]{35};\draw(0.0,48.333333333333336)rectangle(1.6666666666666667,50.0);\draw(0.8333333333333334,50.0)node[anchor=north]{37};\draw(0.8333333333333334,48.333333333333336)node[anchor=south]{36};\draw(0.0,50.0)rectangle(1.6666666666666667,51.66666666666667);\draw(0.8333333333333334,51.66666666666667)node[anchor=north]{38};\draw(0.8333333333333334,50.0)node[anchor=south]{37};\draw(0.0,51.66666666666667)rectangle(1.6666666666666667,53.333333333333336);\draw(0.8333333333333334,53.333333333333336)node[anchor=north]{39};\draw(0.8333333333333334,51.66666666666667)node[anchor=south]{38};\draw(0.0,53.333333333333336)rectangle(1.6666666666666667,55.0);\draw(0.8333333333333334,55.0)node[anchor=north]{40};\draw(0.8333333333333334,53.333333333333336)node[anchor=south]{39};\draw(0.0,55.0)rectangle(1.6666666666666667,56.66666666666667);\draw(0.8333333333333334,56.66666666666667)node[anchor=north]{41};\draw(0.8333333333333334,55.0)node[anchor=south]{40};\draw(0.0,56.66666666666667)rectangle(1.6666666666666667,58.333333333333336);\draw(0.8333333333333334,58.333333333333336)node[anchor=north]{42};\draw(0.8333333333333334,56.66666666666667)node[anchor=south]{41};\draw(0.0,58.333333333333336)rectangle(1.6666666666666667,60.0);\draw(0.8333333333333334,60.0)node[anchor=north]{43};\draw(0.8333333333333334,58.333333333333336)node[anchor=south]{42};\draw(0.0,60.0)rectangle(1.6666666666666667,61.66666666666667);\draw(0.8333333333333334,61.66666666666667)node[anchor=north]{44};\draw(0.8333333333333334,60.0)node[anchor=south]{43};\draw(0.0,61.66666666666667)rectangle(1.6666666666666667,63.333333333333336);\draw(0.8333333333333334,63.333333333333336)node[anchor=north]{45};\draw(0.8333333333333334,61.66666666666667)node[anchor=south]{44};\draw(0.0,63.333333333333336)rectangle(1.6666666666666667,65.0);\draw(1.6666666666666667,64.16666666666667)node[anchor=east]{46};\draw(0.8333333333333334,63.333333333333336)node[anchor=south]{45};\draw(1.6666666666666667,43.333333333333336)rectangle(3.3333333333333335,45.0);\draw(1.6666666666666667,44.16666666666667)node[anchor=west]{33};\draw(3.3333333333333335,44.16666666666667)node[anchor=east]{32};\draw(1.6666666666666667,45.0)rectangle(3.3333333333333335,46.66666666666667);\draw(2.5,46.66666666666667)node[anchor=north]{57};\draw(3.3333333333333335,45.833333333333336)node[anchor=east]{58};\draw(1.6666666666666667,46.66666666666667)rectangle(3.3333333333333335,48.333333333333336);\draw(2.5,48.333333333333336)node[anchor=north]{56};\draw(3.3333333333333335,47.5)node[anchor=east]{63};\draw(2.5,46.66666666666667)node[anchor=south]{57};\draw(1.6666666666666667,48.333333333333336)rectangle(3.3333333333333335,50.0);\draw(2.5,50.0)node[anchor=north]{55};\draw(2.5,48.333333333333336)node[anchor=south]{56};\draw(1.6666666666666667,50.0)rectangle(3.3333333333333335,51.66666666666667);\draw(2.5,51.66666666666667)node[anchor=north]{54};\draw(2.5,50.0)node[anchor=south]{55};\draw(1.6666666666666667,51.66666666666667)rectangle(3.3333333333333335,53.333333333333336);\draw(2.5,53.333333333333336)node[anchor=north]{53};\draw(2.5,51.66666666666667)node[anchor=south]{54};\draw(1.6666666666666667,53.333333333333336)rectangle(3.3333333333333335,55.0);\draw(2.5,55.0)node[anchor=north]{52};\draw(2.5,53.333333333333336)node[anchor=south]{53};\draw(1.6666666666666667,55.0)rectangle(3.3333333333333335,56.66666666666667);\draw(2.5,56.66666666666667)node[anchor=north]{51};\draw(2.5,55.0)node[anchor=south]{52};\draw(1.6666666666666667,56.66666666666667)rectangle(3.3333333333333335,58.333333333333336);\draw(2.5,58.333333333333336)node[anchor=north]{50};\draw(2.5,56.66666666666667)node[anchor=south]{51};\draw(1.6666666666666667,58.333333333333336)rectangle(3.3333333333333335,60.0);\draw(2.5,60.0)node[anchor=north]{49};\draw(2.5,58.333333333333336)node[anchor=south]{50};\draw(1.6666666666666667,60.0)rectangle(3.3333333333333335,61.66666666666667);\draw(2.5,61.66666666666667)node[anchor=north]{48};\draw(2.5,60.0)node[anchor=south]{49};\draw(1.6666666666666667,61.66666666666667)rectangle(3.3333333333333335,63.333333333333336);\draw(2.5,63.333333333333336)node[anchor=north]{47};\draw(2.5,61.66666666666667)node[anchor=south]{48};\draw(1.6666666666666667,63.333333333333336)rectangle(3.3333333333333335,65.0);\draw(1.6666666666666667,64.16666666666667)node[anchor=west]{46};\draw(2.5,63.333333333333336)node[anchor=south]{47};\draw(1.6666666666666667,65.0)rectangle(3.3333333333333335,66.66666666666667);\draw(2.5,66.66666666666667)node[anchor=north]{36};\draw(3.3333333333333335,65.83333333333334)node[anchor=east]{65};\draw(1.6666666666666667,66.66666666666667)rectangle(3.3333333333333335,68.33333333333334);\draw(2.5,68.33333333333334)node[anchor=north]{37};\draw(2.5,66.66666666666667)node[anchor=south]{36};\draw(1.6666666666666667,68.33333333333334)rectangle(3.3333333333333335,70.0);\draw(2.5,70.0)node[anchor=north]{38};\draw(2.5,68.33333333333334)node[anchor=south]{37};\draw(1.6666666666666667,70.0)rectangle(3.3333333333333335,71.66666666666667);\draw(2.5,71.66666666666667)node[anchor=north]{39};\draw(2.5,70.0)node[anchor=south]{38};\draw(1.6666666666666667,71.66666666666667)rectangle(3.3333333333333335,73.33333333333334);\draw(2.5,73.33333333333334)node[anchor=north]{40};\draw(2.5,71.66666666666667)node[anchor=south]{39};\draw(1.6666666666666667,73.33333333333334)rectangle(3.3333333333333335,75.0);\draw(2.5,75.0)node[anchor=north]{41};\draw(2.5,73.33333333333334)node[anchor=south]{40};\draw(1.6666666666666667,75.0)rectangle(3.3333333333333335,76.66666666666667);\draw(2.5,76.66666666666667)node[anchor=north]{42};\draw(2.5,75.0)node[anchor=south]{41};\draw(1.6666666666666667,76.66666666666667)rectangle(3.3333333333333335,78.33333333333334);\draw(2.5,78.33333333333334)node[anchor=north]{43};\draw(2.5,76.66666666666667)node[anchor=south]{42};\draw(1.6666666666666667,78.33333333333334)rectangle(3.3333333333333335,80.0);\draw(2.5,80.0)node[anchor=north]{44};\draw(2.5,78.33333333333334)node[anchor=south]{43};\draw(1.6666666666666667,80.0)rectangle(3.3333333333333335,81.66666666666667);\draw(2.5,81.66666666666667)node[anchor=north]{45};\draw(2.5,80.0)node[anchor=south]{44};\draw(1.6666666666666667,81.66666666666667)rectangle(3.3333333333333335,83.33333333333334);\draw(3.3333333333333335,82.5)node[anchor=east]{46};\draw(2.5,81.66666666666667)node[anchor=south]{45};\draw(3.3333333333333335,43.333333333333336)rectangle(5.0,45.0);\draw(3.3333333333333335,44.16666666666667)node[anchor=west]{32};\draw(5.0,44.16666666666667)node[anchor=east]{31};\draw(3.3333333333333335,45.0)rectangle(5.0,46.66666666666667);\draw(3.3333333333333335,45.833333333333336)node[anchor=west]{58};\draw(5.0,45.833333333333336)node[anchor=east]{59};\draw(3.3333333333333335,46.66666666666667)rectangle(5.0,48.333333333333336);\draw(3.3333333333333335,47.5)node[anchor=west]{63};\draw(4.166666666666667,48.333333333333336)node[anchor=north]{19};\draw(3.3333333333333335,48.333333333333336)rectangle(5.0,50.0);\draw(4.166666666666667,50.0)node[anchor=north]{20};\draw(4.166666666666667,48.333333333333336)node[anchor=south]{19};\draw(3.3333333333333335,50.0)rectangle(5.0,51.66666666666667);\draw(4.166666666666667,51.66666666666667)node[anchor=north]{21};\draw(4.166666666666667,50.0)node[anchor=south]{20};\draw(3.3333333333333335,51.66666666666667)rectangle(5.0,53.333333333333336);\draw(4.166666666666667,53.333333333333336)node[anchor=north]{22};\draw(4.166666666666667,51.66666666666667)node[anchor=south]{21};\draw(3.3333333333333335,53.333333333333336)rectangle(5.0,55.0);\draw(4.166666666666667,55.0)node[anchor=north]{23};\draw(4.166666666666667,53.333333333333336)node[anchor=south]{22};\draw(3.3333333333333335,55.0)rectangle(5.0,56.66666666666667);\draw(4.166666666666667,56.66666666666667)node[anchor=north]{24};\draw(4.166666666666667,55.0)node[anchor=south]{23};\draw(3.3333333333333335,56.66666666666667)rectangle(5.0,58.333333333333336);\draw(4.166666666666667,58.333333333333336)node[anchor=north]{25};\draw(4.166666666666667,56.66666666666667)node[anchor=south]{24};\draw(3.3333333333333335,58.333333333333336)rectangle(5.0,60.0);\draw(4.166666666666667,60.0)node[anchor=north]{26};\draw(4.166666666666667,58.333333333333336)node[anchor=south]{25};\draw(3.3333333333333335,60.0)rectangle(5.0,61.66666666666667);\draw(4.166666666666667,61.66666666666667)node[anchor=north]{27};\draw(4.166666666666667,60.0)node[anchor=south]{26};\draw(3.3333333333333335,61.66666666666667)rectangle(5.0,63.333333333333336);\draw(4.166666666666667,63.333333333333336)node[anchor=north]{28};\draw(4.166666666666667,61.66666666666667)node[anchor=south]{27};\draw(3.3333333333333335,63.333333333333336)rectangle(5.0,65.0);\draw(3.3333333333333335,64.16666666666667)node[anchor=west]{29};\draw(4.166666666666667,65.0)node[anchor=north]{64};\draw(4.166666666666667,63.333333333333336)node[anchor=south]{28};\draw(3.3333333333333335,65.0)rectangle(5.0,66.66666666666667);\draw(3.3333333333333335,65.83333333333334)node[anchor=west]{65};\draw(4.166666666666667,65.0)node[anchor=south]{64};\draw(3.3333333333333335,66.66666666666667)rectangle(5.0,68.33333333333334);\draw(4.166666666666667,68.33333333333334)node[anchor=north]{55};\draw(4.166666666666667,66.66666666666667)node[anchor=south]{56};\draw(3.3333333333333335,68.33333333333334)rectangle(5.0,70.0);\draw(4.166666666666667,70.0)node[anchor=north]{54};\draw(4.166666666666667,68.33333333333334)node[anchor=south]{55};\draw(3.3333333333333335,70.0)rectangle(5.0,71.66666666666667);\draw(4.166666666666667,71.66666666666667)node[anchor=north]{53};\draw(4.166666666666667,70.0)node[anchor=south]{54};\draw(3.3333333333333335,71.66666666666667)rectangle(5.0,73.33333333333334);\draw(4.166666666666667,73.33333333333334)node[anchor=north]{52};\draw(4.166666666666667,71.66666666666667)node[anchor=south]{53};\draw(3.3333333333333335,73.33333333333334)rectangle(5.0,75.0);\draw(4.166666666666667,75.0)node[anchor=north]{51};\draw(4.166666666666667,73.33333333333334)node[anchor=south]{52};\draw(3.3333333333333335,75.0)rectangle(5.0,76.66666666666667);\draw(4.166666666666667,76.66666666666667)node[anchor=north]{50};\draw(4.166666666666667,75.0)node[anchor=south]{51};\draw(3.3333333333333335,76.66666666666667)rectangle(5.0,78.33333333333334);\draw(4.166666666666667,78.33333333333334)node[anchor=north]{49};\draw(4.166666666666667,76.66666666666667)node[anchor=south]{50};\draw(3.3333333333333335,78.33333333333334)rectangle(5.0,80.0);\draw(4.166666666666667,80.0)node[anchor=north]{48};\draw(4.166666666666667,78.33333333333334)node[anchor=south]{49};\draw(3.3333333333333335,80.0)rectangle(5.0,81.66666666666667);\draw(4.166666666666667,81.66666666666667)node[anchor=north]{47};\draw(4.166666666666667,80.0)node[anchor=south]{48};\draw(3.3333333333333335,81.66666666666667)rectangle(5.0,83.33333333333334);\draw(3.3333333333333335,82.5)node[anchor=west]{46};\draw(4.166666666666667,81.66666666666667)node[anchor=south]{47};\draw(5.0,43.333333333333336)rectangle(6.666666666666667,45.0);\draw(5.0,44.16666666666667)node[anchor=west]{31};\draw(6.666666666666667,44.16666666666667)node[anchor=east]{30};\draw(5.0,45.0)rectangle(6.666666666666667,46.66666666666667);\draw(5.0,45.833333333333336)node[anchor=west]{59};\draw(5.833333333333334,46.66666666666667)node[anchor=north]{3};\draw(5.0,46.66666666666667)rectangle(6.666666666666667,48.333333333333336);\draw(5.833333333333334,48.333333333333336)node[anchor=north]{4};\draw(5.833333333333334,46.66666666666667)node[anchor=south]{3};\draw(5.0,48.333333333333336)rectangle(6.666666666666667,50.0);\draw(5.833333333333334,50.0)node[anchor=north]{5};\draw(5.833333333333334,48.333333333333336)node[anchor=south]{4};\draw(5.0,50.0)rectangle(6.666666666666667,51.66666666666667);\draw(5.833333333333334,51.66666666666667)node[anchor=north]{6};\draw(5.833333333333334,50.0)node[anchor=south]{5};\draw(5.0,51.66666666666667)rectangle(6.666666666666667,53.333333333333336);\draw(5.833333333333334,53.333333333333336)node[anchor=north]{7};\draw(5.833333333333334,51.66666666666667)node[anchor=south]{6};\draw(5.0,53.333333333333336)rectangle(6.666666666666667,55.0);\draw(5.833333333333334,55.0)node[anchor=north]{8};\draw(5.833333333333334,53.333333333333336)node[anchor=south]{7};\draw(5.0,55.0)rectangle(6.666666666666667,56.66666666666667);\draw(5.833333333333334,56.66666666666667)node[anchor=north]{9};\draw(5.833333333333334,55.0)node[anchor=south]{8};\draw(5.0,56.66666666666667)rectangle(6.666666666666667,58.333333333333336);\draw(5.833333333333334,58.333333333333336)node[anchor=north]{10};\draw(5.833333333333334,56.66666666666667)node[anchor=south]{9};\draw(5.0,58.333333333333336)rectangle(6.666666666666667,60.0);\draw(5.833333333333334,60.0)node[anchor=north]{11};\draw(5.833333333333334,58.333333333333336)node[anchor=south]{10};\draw(5.0,60.0)rectangle(6.666666666666667,61.66666666666667);\draw(5.833333333333334,61.66666666666667)node[anchor=north]{12};\draw(5.833333333333334,60.0)node[anchor=south]{11};\draw(5.0,61.66666666666667)rectangle(6.666666666666667,63.333333333333336);\draw(5.833333333333334,63.333333333333336)node[anchor=north]{13};\draw(5.833333333333334,61.66666666666667)node[anchor=south]{12};\draw(5.0,63.333333333333336)rectangle(6.666666666666667,65.0);\draw(5.0,64.16666666666667)node[anchor=west]{14};\draw(6.666666666666667,64.16666666666667)node[anchor=east]{60};\draw(5.833333333333334,63.333333333333336)node[anchor=south]{13};\draw(5.0,65.0)rectangle(6.666666666666667,66.66666666666667);\draw(5.833333333333334,66.66666666666667)node[anchor=north]{35};\draw(6.666666666666667,65.83333333333334)node[anchor=east]{62};\draw(5.0,66.66666666666667)rectangle(6.666666666666667,68.33333333333334);\draw(5.833333333333334,68.33333333333334)node[anchor=north]{36};\draw(5.833333333333334,66.66666666666667)node[anchor=south]{35};\draw(5.0,68.33333333333334)rectangle(6.666666666666667,70.0);\draw(5.833333333333334,70.0)node[anchor=north]{37};\draw(5.833333333333334,68.33333333333334)node[anchor=south]{36};\draw(5.0,70.0)rectangle(6.666666666666667,71.66666666666667);\draw(5.833333333333334,71.66666666666667)node[anchor=north]{38};\draw(5.833333333333334,70.0)node[anchor=south]{37};\draw(5.0,71.66666666666667)rectangle(6.666666666666667,73.33333333333334);\draw(5.833333333333334,73.33333333333334)node[anchor=north]{39};\draw(5.833333333333334,71.66666666666667)node[anchor=south]{38};\draw(5.0,73.33333333333334)rectangle(6.666666666666667,75.0);\draw(5.833333333333334,75.0)node[anchor=north]{40};\draw(5.833333333333334,73.33333333333334)node[anchor=south]{39};\draw(5.0,75.0)rectangle(6.666666666666667,76.66666666666667);\draw(5.833333333333334,76.66666666666667)node[anchor=north]{41};\draw(5.833333333333334,75.0)node[anchor=south]{40};\draw(5.0,76.66666666666667)rectangle(6.666666666666667,78.33333333333334);\draw(5.833333333333334,78.33333333333334)node[anchor=north]{42};\draw(5.833333333333334,76.66666666666667)node[anchor=south]{41};\draw(5.0,78.33333333333334)rectangle(6.666666666666667,80.0);\draw(5.833333333333334,80.0)node[anchor=north]{43};\draw(5.833333333333334,78.33333333333334)node[anchor=south]{42};\draw(5.0,80.0)rectangle(6.666666666666667,81.66666666666667);\draw(5.833333333333334,81.66666666666667)node[anchor=north]{44};\draw(5.833333333333334,80.0)node[anchor=south]{43};\draw(5.0,81.66666666666667)rectangle(6.666666666666667,83.33333333333334);\draw(5.833333333333334,83.33333333333334)node[anchor=north]{45};\draw(5.833333333333334,81.66666666666667)node[anchor=south]{44};\draw(5.0,83.33333333333334)rectangle(6.666666666666667,85.0);\draw(6.666666666666667,84.16666666666667)node[anchor=east]{46};\draw(5.833333333333334,83.33333333333334)node[anchor=south]{45};\draw(6.666666666666667,43.333333333333336)rectangle(8.333333333333334,45.0);\draw(6.666666666666667,44.16666666666667)node[anchor=west]{30};\draw(8.333333333333334,44.16666666666667)node[anchor=east]{29};\draw(6.666666666666667,45.0)rectangle(8.333333333333334,46.66666666666667);\draw(7.5,46.66666666666667)node[anchor=north]{36};\draw(8.333333333333334,45.833333333333336)node[anchor=east]{65};\draw(6.666666666666667,46.66666666666667)rectangle(8.333333333333334,48.333333333333336);\draw(7.5,48.333333333333336)node[anchor=north]{37};\draw(7.5,46.66666666666667)node[anchor=south]{36};\draw(6.666666666666667,48.333333333333336)rectangle(8.333333333333334,50.0);\draw(7.5,50.0)node[anchor=north]{38};\draw(7.5,48.333333333333336)node[anchor=south]{37};\draw(6.666666666666667,50.0)rectangle(8.333333333333334,51.66666666666667);\draw(7.5,51.66666666666667)node[anchor=north]{39};\draw(7.5,50.0)node[anchor=south]{38};\draw(6.666666666666667,51.66666666666667)rectangle(8.333333333333334,53.333333333333336);\draw(7.5,53.333333333333336)node[anchor=north]{40};\draw(7.5,51.66666666666667)node[anchor=south]{39};\draw(6.666666666666667,53.333333333333336)rectangle(8.333333333333334,55.0);\draw(7.5,55.0)node[anchor=north]{41};\draw(7.5,53.333333333333336)node[anchor=south]{40};\draw(6.666666666666667,55.0)rectangle(8.333333333333334,56.66666666666667);\draw(7.5,56.66666666666667)node[anchor=north]{42};\draw(7.5,55.0)node[anchor=south]{41};\draw(6.666666666666667,56.66666666666667)rectangle(8.333333333333334,58.333333333333336);\draw(7.5,58.333333333333336)node[anchor=north]{43};\draw(7.5,56.66666666666667)node[anchor=south]{42};\draw(6.666666666666667,58.333333333333336)rectangle(8.333333333333334,60.0);\draw(7.5,60.0)node[anchor=north]{44};\draw(7.5,58.333333333333336)node[anchor=south]{43};\draw(6.666666666666667,60.0)rectangle(8.333333333333334,61.66666666666667);\draw(7.5,61.66666666666667)node[anchor=north]{45};\draw(7.5,60.0)node[anchor=south]{44};\draw(6.666666666666667,61.66666666666667)rectangle(8.333333333333334,63.333333333333336);\draw(8.333333333333334,62.5)node[anchor=east]{46};\draw(7.5,61.66666666666667)node[anchor=south]{45};\draw(6.666666666666667,63.333333333333336)rectangle(8.333333333333334,65.0);\draw(6.666666666666667,64.16666666666667)node[anchor=west]{60};\draw(7.5,65.0)node[anchor=north]{61};\draw(6.666666666666667,65.0)rectangle(8.333333333333334,66.66666666666667);\draw(6.666666666666667,65.83333333333334)node[anchor=west]{62};\draw(7.5,65.0)node[anchor=south]{61};\draw(6.666666666666667,66.66666666666667)rectangle(8.333333333333334,68.33333333333334);\draw(7.5,68.33333333333334)node[anchor=north]{56};\draw(8.333333333333334,67.5)node[anchor=east]{63};\draw(7.5,66.66666666666667)node[anchor=south]{57};\draw(6.666666666666667,68.33333333333334)rectangle(8.333333333333334,70.0);\draw(7.5,70.0)node[anchor=north]{55};\draw(7.5,68.33333333333334)node[anchor=south]{56};\draw(6.666666666666667,70.0)rectangle(8.333333333333334,71.66666666666667);\draw(7.5,71.66666666666667)node[anchor=north]{54};\draw(7.5,70.0)node[anchor=south]{55};\draw(6.666666666666667,71.66666666666667)rectangle(8.333333333333334,73.33333333333334);\draw(7.5,73.33333333333334)node[anchor=north]{53};\draw(7.5,71.66666666666667)node[anchor=south]{54};\draw(6.666666666666667,73.33333333333334)rectangle(8.333333333333334,75.0);\draw(7.5,75.0)node[anchor=north]{52};\draw(7.5,73.33333333333334)node[anchor=south]{53};\draw(6.666666666666667,75.0)rectangle(8.333333333333334,76.66666666666667);\draw(7.5,76.66666666666667)node[anchor=north]{51};\draw(7.5,75.0)node[anchor=south]{52};\draw(6.666666666666667,76.66666666666667)rectangle(8.333333333333334,78.33333333333334);\draw(7.5,78.33333333333334)node[anchor=north]{50};\draw(7.5,76.66666666666667)node[anchor=south]{51};\draw(6.666666666666667,78.33333333333334)rectangle(8.333333333333334,80.0);\draw(7.5,80.0)node[anchor=north]{49};\draw(7.5,78.33333333333334)node[anchor=south]{50};\draw(6.666666666666667,80.0)rectangle(8.333333333333334,81.66666666666667);\draw(7.5,81.66666666666667)node[anchor=north]{48};\draw(7.5,80.0)node[anchor=south]{49};\draw(6.666666666666667,81.66666666666667)rectangle(8.333333333333334,83.33333333333334);\draw(7.5,83.33333333333334)node[anchor=north]{47};\draw(7.5,81.66666666666667)node[anchor=south]{48};\draw(6.666666666666667,83.33333333333334)rectangle(8.333333333333334,85.0);\draw(6.666666666666667,84.16666666666667)node[anchor=west]{46};\draw(7.5,83.33333333333334)node[anchor=south]{47};\draw(6.666666666666667,85.0)rectangle(8.333333333333334,86.66666666666667);\draw(7.5,86.66666666666667)node[anchor=north]{36};\draw(8.333333333333334,85.83333333333334)node[anchor=east]{65};\draw(6.666666666666667,86.66666666666667)rectangle(8.333333333333334,88.33333333333334);\draw(7.5,88.33333333333334)node[anchor=north]{37};\draw(7.5,86.66666666666667)node[anchor=south]{36};\draw(6.666666666666667,88.33333333333334)rectangle(8.333333333333334,90.0);\draw(7.5,90.0)node[anchor=north]{38};\draw(7.5,88.33333333333334)node[anchor=south]{37};\draw(6.666666666666667,90.0)rectangle(8.333333333333334,91.66666666666667);\draw(7.5,91.66666666666667)node[anchor=north]{39};\draw(7.5,90.0)node[anchor=south]{38};\draw(6.666666666666667,91.66666666666667)rectangle(8.333333333333334,93.33333333333334);\draw(7.5,93.33333333333334)node[anchor=north]{40};\draw(7.5,91.66666666666667)node[anchor=south]{39};\draw(6.666666666666667,93.33333333333334)rectangle(8.333333333333334,95.0);\draw(7.5,95.0)node[anchor=north]{41};\draw(7.5,93.33333333333334)node[anchor=south]{40};\draw(6.666666666666667,95.0)rectangle(8.333333333333334,96.66666666666667);\draw(7.5,96.66666666666667)node[anchor=north]{42};\draw(7.5,95.0)node[anchor=south]{41};\draw(6.666666666666667,96.66666666666667)rectangle(8.333333333333334,98.33333333333334);\draw(7.5,98.33333333333334)node[anchor=north]{43};\draw(7.5,96.66666666666667)node[anchor=south]{42};\draw(6.666666666666667,98.33333333333334)rectangle(8.333333333333334,100.0);\draw(7.5,100.0)node[anchor=north]{44};\draw(7.5,98.33333333333334)node[anchor=south]{43};\draw(6.666666666666667,100.0)rectangle(8.333333333333334,101.66666666666667);\draw(7.5,101.66666666666667)node[anchor=north]{45};\draw(7.5,100.0)node[anchor=south]{44};\draw(6.666666666666667,101.66666666666667)rectangle(8.333333333333334,103.33333333333334);\draw(8.333333333333334,102.5)node[anchor=east]{46};\draw(7.5,101.66666666666667)node[anchor=south]{45};\draw(8.333333333333334,21.666666666666668)rectangle(10.0,23.333333333333336);\draw(9.166666666666668,23.333333333333336)node[anchor=north]{16};\draw(10.0,22.5)node[anchor=east]{15};\draw(8.333333333333334,23.333333333333336)rectangle(10.0,25.0);\draw(9.166666666666668,25.0)node[anchor=north]{17};\draw(9.166666666666668,23.333333333333336)node[anchor=south]{16};\draw(8.333333333333334,25.0)rectangle(10.0,26.666666666666668);\draw(9.166666666666668,26.666666666666668)node[anchor=north]{18};\draw(9.166666666666668,25.0)node[anchor=south]{17};\draw(8.333333333333334,26.666666666666668)rectangle(10.0,28.333333333333336);\draw(9.166666666666668,28.333333333333336)node[anchor=north]{19};\draw(9.166666666666668,26.666666666666668)node[anchor=south]{18};\draw(8.333333333333334,28.333333333333336)rectangle(10.0,30.0);\draw(9.166666666666668,30.0)node[anchor=north]{20};\draw(9.166666666666668,28.333333333333336)node[anchor=south]{19};\draw(8.333333333333334,30.0)rectangle(10.0,31.666666666666668);\draw(9.166666666666668,31.666666666666668)node[anchor=north]{21};\draw(9.166666666666668,30.0)node[anchor=south]{20};\draw(8.333333333333334,31.666666666666668)rectangle(10.0,33.333333333333336);\draw(9.166666666666668,33.333333333333336)node[anchor=north]{22};\draw(9.166666666666668,31.666666666666668)node[anchor=south]{21};\draw(8.333333333333334,33.333333333333336)rectangle(10.0,35.0);\draw(9.166666666666668,35.0)node[anchor=north]{23};\draw(9.166666666666668,33.333333333333336)node[anchor=south]{22};\draw(8.333333333333334,35.0)rectangle(10.0,36.66666666666667);\draw(9.166666666666668,36.66666666666667)node[anchor=north]{24};\draw(9.166666666666668,35.0)node[anchor=south]{23};\draw(8.333333333333334,36.66666666666667)rectangle(10.0,38.333333333333336);\draw(9.166666666666668,38.333333333333336)node[anchor=north]{25};\draw(9.166666666666668,36.66666666666667)node[anchor=south]{24};\draw(8.333333333333334,38.333333333333336)rectangle(10.0,40.0);\draw(9.166666666666668,40.0)node[anchor=north]{26};\draw(9.166666666666668,38.333333333333336)node[anchor=south]{25};\draw(8.333333333333334,40.0)rectangle(10.0,41.66666666666667);\draw(9.166666666666668,41.66666666666667)node[anchor=north]{27};\draw(9.166666666666668,40.0)node[anchor=south]{26};\draw(8.333333333333334,41.66666666666667)rectangle(10.0,43.333333333333336);\draw(9.166666666666668,43.333333333333336)node[anchor=north]{28};\draw(9.166666666666668,41.66666666666667)node[anchor=south]{27};\draw(8.333333333333334,43.333333333333336)rectangle(10.0,45.0);\draw(8.333333333333334,44.16666666666667)node[anchor=west]{29};\draw(9.166666666666668,45.0)node[anchor=north]{64};\draw(9.166666666666668,43.333333333333336)node[anchor=south]{28};\draw(8.333333333333334,45.0)rectangle(10.0,46.66666666666667);\draw(8.333333333333334,45.833333333333336)node[anchor=west]{65};\draw(9.166666666666668,45.0)node[anchor=south]{64};\draw(8.333333333333334,46.66666666666667)rectangle(10.0,48.333333333333336);\draw(9.166666666666668,48.333333333333336)node[anchor=north]{55};\draw(9.166666666666668,46.66666666666667)node[anchor=south]{56};\draw(8.333333333333334,48.333333333333336)rectangle(10.0,50.0);\draw(9.166666666666668,50.0)node[anchor=north]{54};\draw(9.166666666666668,48.333333333333336)node[anchor=south]{55};\draw(8.333333333333334,50.0)rectangle(10.0,51.66666666666667);\draw(9.166666666666668,51.66666666666667)node[anchor=north]{53};\draw(9.166666666666668,50.0)node[anchor=south]{54};\draw(8.333333333333334,51.66666666666667)rectangle(10.0,53.333333333333336);\draw(9.166666666666668,53.333333333333336)node[anchor=north]{52};\draw(9.166666666666668,51.66666666666667)node[anchor=south]{53};\draw(8.333333333333334,53.333333333333336)rectangle(10.0,55.0);\draw(9.166666666666668,55.0)node[anchor=north]{51};\draw(9.166666666666668,53.333333333333336)node[anchor=south]{52};\draw(8.333333333333334,55.0)rectangle(10.0,56.66666666666667);\draw(9.166666666666668,56.66666666666667)node[anchor=north]{50};\draw(9.166666666666668,55.0)node[anchor=south]{51};\draw(8.333333333333334,56.66666666666667)rectangle(10.0,58.333333333333336);\draw(9.166666666666668,58.333333333333336)node[anchor=north]{49};\draw(9.166666666666668,56.66666666666667)node[anchor=south]{50};\draw(8.333333333333334,58.333333333333336)rectangle(10.0,60.0);\draw(9.166666666666668,60.0)node[anchor=north]{48};\draw(9.166666666666668,58.333333333333336)node[anchor=south]{49};\draw(8.333333333333334,60.0)rectangle(10.0,61.66666666666667);\draw(9.166666666666668,61.66666666666667)node[anchor=north]{47};\draw(9.166666666666668,60.0)node[anchor=south]{48};\draw(8.333333333333334,61.66666666666667)rectangle(10.0,63.333333333333336);\draw(8.333333333333334,62.5)node[anchor=west]{46};\draw(9.166666666666668,61.66666666666667)node[anchor=south]{47};\draw(8.333333333333334,66.66666666666667)rectangle(10.0,68.33333333333334);\draw(8.333333333333334,67.5)node[anchor=west]{63};\draw(9.166666666666668,68.33333333333334)node[anchor=north]{19};\draw(8.333333333333334,68.33333333333334)rectangle(10.0,70.0);\draw(9.166666666666668,70.0)node[anchor=north]{20};\draw(9.166666666666668,68.33333333333334)node[anchor=south]{19};\draw(8.333333333333334,70.0)rectangle(10.0,71.66666666666667);\draw(9.166666666666668,71.66666666666667)node[anchor=north]{21};\draw(9.166666666666668,70.0)node[anchor=south]{20};\draw(8.333333333333334,71.66666666666667)rectangle(10.0,73.33333333333334);\draw(9.166666666666668,73.33333333333334)node[anchor=north]{22};\draw(9.166666666666668,71.66666666666667)node[anchor=south]{21};\draw(8.333333333333334,73.33333333333334)rectangle(10.0,75.0);\draw(9.166666666666668,75.0)node[anchor=north]{23};\draw(9.166666666666668,73.33333333333334)node[anchor=south]{22};\draw(8.333333333333334,75.0)rectangle(10.0,76.66666666666667);\draw(9.166666666666668,76.66666666666667)node[anchor=north]{24};\draw(9.166666666666668,75.0)node[anchor=south]{23};\draw(8.333333333333334,76.66666666666667)rectangle(10.0,78.33333333333334);\draw(9.166666666666668,78.33333333333334)node[anchor=north]{25};\draw(9.166666666666668,76.66666666666667)node[anchor=south]{24};\draw(8.333333333333334,78.33333333333334)rectangle(10.0,80.0);\draw(9.166666666666668,80.0)node[anchor=north]{26};\draw(9.166666666666668,78.33333333333334)node[anchor=south]{25};\draw(8.333333333333334,80.0)rectangle(10.0,81.66666666666667);\draw(9.166666666666668,81.66666666666667)node[anchor=north]{27};\draw(9.166666666666668,80.0)node[anchor=south]{26};\draw(8.333333333333334,81.66666666666667)rectangle(10.0,83.33333333333334);\draw(9.166666666666668,83.33333333333334)node[anchor=north]{28};\draw(9.166666666666668,81.66666666666667)node[anchor=south]{27};\draw(8.333333333333334,83.33333333333334)rectangle(10.0,85.0);\draw(8.333333333333334,84.16666666666667)node[anchor=west]{29};\draw(9.166666666666668,85.0)node[anchor=north]{64};\draw(9.166666666666668,83.33333333333334)node[anchor=south]{28};\draw(8.333333333333334,85.0)rectangle(10.0,86.66666666666667);\draw(8.333333333333334,85.83333333333334)node[anchor=west]{65};\draw(9.166666666666668,85.0)node[anchor=south]{64};\draw(8.333333333333334,86.66666666666667)rectangle(10.0,88.33333333333334);\draw(9.166666666666668,88.33333333333334)node[anchor=north]{55};\draw(9.166666666666668,86.66666666666667)node[anchor=south]{56};\draw(8.333333333333334,88.33333333333334)rectangle(10.0,90.0);\draw(9.166666666666668,90.0)node[anchor=north]{54};\draw(9.166666666666668,88.33333333333334)node[anchor=south]{55};\draw(8.333333333333334,90.0)rectangle(10.0,91.66666666666667);\draw(9.166666666666668,91.66666666666667)node[anchor=north]{53};\draw(9.166666666666668,90.0)node[anchor=south]{54};\draw(8.333333333333334,91.66666666666667)rectangle(10.0,93.33333333333334);\draw(9.166666666666668,93.33333333333334)node[anchor=north]{52};\draw(9.166666666666668,91.66666666666667)node[anchor=south]{53};\draw(8.333333333333334,93.33333333333334)rectangle(10.0,95.0);\draw(9.166666666666668,95.0)node[anchor=north]{51};\draw(9.166666666666668,93.33333333333334)node[anchor=south]{52};\draw(8.333333333333334,95.0)rectangle(10.0,96.66666666666667);\draw(9.166666666666668,96.66666666666667)node[anchor=north]{50};\draw(9.166666666666668,95.0)node[anchor=south]{51};\draw(8.333333333333334,96.66666666666667)rectangle(10.0,98.33333333333334);\draw(9.166666666666668,98.33333333333334)node[anchor=north]{49};\draw(9.166666666666668,96.66666666666667)node[anchor=south]{50};\draw(8.333333333333334,98.33333333333334)rectangle(10.0,100.0);\draw(9.166666666666668,100.0)node[anchor=north]{48};\draw(9.166666666666668,98.33333333333334)node[anchor=south]{49};\draw(8.333333333333334,100.0)rectangle(10.0,101.66666666666667);\draw(9.166666666666668,101.66666666666667)node[anchor=north]{47};\draw(9.166666666666668,100.0)node[anchor=south]{48};\draw(8.333333333333334,101.66666666666667)rectangle(10.0,103.33333333333334);\draw(8.333333333333334,102.5)node[anchor=west]{46};\draw(9.166666666666668,101.66666666666667)node[anchor=south]{47};\draw(10.0,21.666666666666668)rectangle(11.666666666666668,23.333333333333336);\draw(10.0,22.5)node[anchor=west]{15};\draw(11.666666666666668,22.5)node[anchor=east]{14};\draw(11.666666666666668,0.0)rectangle(13.333333333333334,1.6666666666666667);\draw(12.5,1.6666666666666667)node[anchor=north]{1};\draw(11.666666666666668,1.6666666666666667)rectangle(13.333333333333334,3.3333333333333335);\draw(12.5,3.3333333333333335)node[anchor=north]{2};\draw(12.5,1.6666666666666667)node[anchor=south]{1};\draw(11.666666666666668,3.3333333333333335)rectangle(13.333333333333334,5.0);\draw(12.5,5.0)node[anchor=north]{3};\draw(12.5,3.3333333333333335)node[anchor=south]{2};\draw(11.666666666666668,5.0)rectangle(13.333333333333334,6.666666666666667);\draw(12.5,6.666666666666667)node[anchor=north]{4};\draw(12.5,5.0)node[anchor=south]{3};\draw(11.666666666666668,6.666666666666667)rectangle(13.333333333333334,8.333333333333334);\draw(12.5,8.333333333333334)node[anchor=north]{5};\draw(12.5,6.666666666666667)node[anchor=south]{4};\draw(11.666666666666668,8.333333333333334)rectangle(13.333333333333334,10.0);\draw(12.5,10.0)node[anchor=north]{6};\draw(12.5,8.333333333333334)node[anchor=south]{5};\draw(11.666666666666668,10.0)rectangle(13.333333333333334,11.666666666666668);\draw(12.5,11.666666666666668)node[anchor=north]{7};\draw(12.5,10.0)node[anchor=south]{6};\draw(11.666666666666668,11.666666666666668)rectangle(13.333333333333334,13.333333333333334);\draw(12.5,13.333333333333334)node[anchor=north]{8};\draw(12.5,11.666666666666668)node[anchor=south]{7};\draw(11.666666666666668,13.333333333333334)rectangle(13.333333333333334,15.0);\draw(12.5,15.0)node[anchor=north]{9};\draw(12.5,13.333333333333334)node[anchor=south]{8};\draw(11.666666666666668,15.0)rectangle(13.333333333333334,16.666666666666668);\draw(12.5,16.666666666666668)node[anchor=north]{10};\draw(12.5,15.0)node[anchor=south]{9};\draw(11.666666666666668,16.666666666666668)rectangle(13.333333333333334,18.333333333333336);\draw(12.5,18.333333333333336)node[anchor=north]{11};\draw(12.5,16.666666666666668)node[anchor=south]{10};\draw(11.666666666666668,18.333333333333336)rectangle(13.333333333333334,20.0);\draw(12.5,20.0)node[anchor=north]{12};\draw(12.5,18.333333333333336)node[anchor=south]{11};\draw(11.666666666666668,20.0)rectangle(13.333333333333334,21.666666666666668);\draw(12.5,21.666666666666668)node[anchor=north]{13};\draw(12.5,20.0)node[anchor=south]{12};\draw(11.666666666666668,21.666666666666668)rectangle(13.333333333333334,23.333333333333336);\draw(11.666666666666668,22.5)node[anchor=west]{14};\draw(13.333333333333334,22.5)node[anchor=east]{60};\draw(12.5,21.666666666666668)node[anchor=south]{13};\draw(11.666666666666668,23.333333333333336)rectangle(13.333333333333334,25.0);\draw(12.5,25.0)node[anchor=north]{35};\draw(13.333333333333334,24.166666666666668)node[anchor=east]{62};\draw(11.666666666666668,25.0)rectangle(13.333333333333334,26.666666666666668);\draw(12.5,26.666666666666668)node[anchor=north]{36};\draw(12.5,25.0)node[anchor=south]{35};\draw(11.666666666666668,26.666666666666668)rectangle(13.333333333333334,28.333333333333336);\draw(12.5,28.333333333333336)node[anchor=north]{37};\draw(12.5,26.666666666666668)node[anchor=south]{36};\draw(11.666666666666668,28.333333333333336)rectangle(13.333333333333334,30.0);\draw(12.5,30.0)node[anchor=north]{38};\draw(12.5,28.333333333333336)node[anchor=south]{37};\draw(11.666666666666668,30.0)rectangle(13.333333333333334,31.666666666666668);\draw(12.5,31.666666666666668)node[anchor=north]{39};\draw(12.5,30.0)node[anchor=south]{38};\draw(11.666666666666668,31.666666666666668)rectangle(13.333333333333334,33.333333333333336);\draw(12.5,33.333333333333336)node[anchor=north]{40};\draw(12.5,31.666666666666668)node[anchor=south]{39};\draw(11.666666666666668,33.333333333333336)rectangle(13.333333333333334,35.0);\draw(12.5,35.0)node[anchor=north]{41};\draw(12.5,33.333333333333336)node[anchor=south]{40};\draw(11.666666666666668,35.0)rectangle(13.333333333333334,36.66666666666667);\draw(12.5,36.66666666666667)node[anchor=north]{42};\draw(12.5,35.0)node[anchor=south]{41};\draw(11.666666666666668,36.66666666666667)rectangle(13.333333333333334,38.333333333333336);\draw(12.5,38.333333333333336)node[anchor=north]{43};\draw(12.5,36.66666666666667)node[anchor=south]{42};\draw(11.666666666666668,38.333333333333336)rectangle(13.333333333333334,40.0);\draw(12.5,40.0)node[anchor=north]{44};\draw(12.5,38.333333333333336)node[anchor=south]{43};\draw(11.666666666666668,40.0)rectangle(13.333333333333334,41.66666666666667);\draw(12.5,41.66666666666667)node[anchor=north]{45};\draw(12.5,40.0)node[anchor=south]{44};\draw(11.666666666666668,41.66666666666667)rectangle(13.333333333333334,43.333333333333336);\draw(13.333333333333334,42.5)node[anchor=east]{46};\draw(12.5,41.66666666666667)node[anchor=south]{45};\draw(13.333333333333334,21.666666666666668)rectangle(15.0,23.333333333333336);\draw(13.333333333333334,22.5)node[anchor=west]{60};\draw(14.166666666666668,23.333333333333336)node[anchor=north]{61};\draw(13.333333333333334,23.333333333333336)rectangle(15.0,25.0);\draw(13.333333333333334,24.166666666666668)node[anchor=west]{62};\draw(14.166666666666668,23.333333333333336)node[anchor=south]{61};\draw(13.333333333333334,25.0)rectangle(15.0,26.666666666666668);\draw(14.166666666666668,26.666666666666668)node[anchor=north]{56};\draw(15.0,25.833333333333336)node[anchor=east]{63};\draw(14.166666666666668,25.0)node[anchor=south]{57};\draw(13.333333333333334,26.666666666666668)rectangle(15.0,28.333333333333336);\draw(14.166666666666668,28.333333333333336)node[anchor=north]{55};\draw(14.166666666666668,26.666666666666668)node[anchor=south]{56};\draw(13.333333333333334,28.333333333333336)rectangle(15.0,30.0);\draw(14.166666666666668,30.0)node[anchor=north]{54};\draw(14.166666666666668,28.333333333333336)node[anchor=south]{55};\draw(13.333333333333334,30.0)rectangle(15.0,31.666666666666668);\draw(14.166666666666668,31.666666666666668)node[anchor=north]{53};\draw(14.166666666666668,30.0)node[anchor=south]{54};\draw(13.333333333333334,31.666666666666668)rectangle(15.0,33.333333333333336);\draw(14.166666666666668,33.333333333333336)node[anchor=north]{52};\draw(14.166666666666668,31.666666666666668)node[anchor=south]{53};\draw(13.333333333333334,33.333333333333336)rectangle(15.0,35.0);\draw(14.166666666666668,35.0)node[anchor=north]{51};\draw(14.166666666666668,33.333333333333336)node[anchor=south]{52};\draw(13.333333333333334,35.0)rectangle(15.0,36.66666666666667);\draw(14.166666666666668,36.66666666666667)node[anchor=north]{50};\draw(14.166666666666668,35.0)node[anchor=south]{51};\draw(13.333333333333334,36.66666666666667)rectangle(15.0,38.333333333333336);\draw(14.166666666666668,38.333333333333336)node[anchor=north]{49};\draw(14.166666666666668,36.66666666666667)node[anchor=south]{50};\draw(13.333333333333334,38.333333333333336)rectangle(15.0,40.0);\draw(14.166666666666668,40.0)node[anchor=north]{48};\draw(14.166666666666668,38.333333333333336)node[anchor=south]{49};\draw(13.333333333333334,40.0)rectangle(15.0,41.66666666666667);\draw(14.166666666666668,41.66666666666667)node[anchor=north]{47};\draw(14.166666666666668,40.0)node[anchor=south]{48};\draw(13.333333333333334,41.66666666666667)rectangle(15.0,43.333333333333336);\draw(13.333333333333334,42.5)node[anchor=west]{46};\draw(14.166666666666668,41.66666666666667)node[anchor=south]{47};\draw(13.333333333333334,43.333333333333336)rectangle(15.0,45.0);\draw(14.166666666666668,45.0)node[anchor=north]{36};\draw(15.0,44.16666666666667)node[anchor=east]{65};\draw(13.333333333333334,45.0)rectangle(15.0,46.66666666666667);\draw(14.166666666666668,46.66666666666667)node[anchor=north]{37};\draw(14.166666666666668,45.0)node[anchor=south]{36};\draw(13.333333333333334,46.66666666666667)rectangle(15.0,48.333333333333336);\draw(14.166666666666668,48.333333333333336)node[anchor=north]{38};\draw(14.166666666666668,46.66666666666667)node[anchor=south]{37};\draw(13.333333333333334,48.333333333333336)rectangle(15.0,50.0);\draw(14.166666666666668,50.0)node[anchor=north]{39};\draw(14.166666666666668,48.333333333333336)node[anchor=south]{38};\draw(13.333333333333334,50.0)rectangle(15.0,51.66666666666667);\draw(14.166666666666668,51.66666666666667)node[anchor=north]{40};\draw(14.166666666666668,50.0)node[anchor=south]{39};\draw(13.333333333333334,51.66666666666667)rectangle(15.0,53.333333333333336);\draw(14.166666666666668,53.333333333333336)node[anchor=north]{41};\draw(14.166666666666668,51.66666666666667)node[anchor=south]{40};\draw(13.333333333333334,53.333333333333336)rectangle(15.0,55.0);\draw(14.166666666666668,55.0)node[anchor=north]{42};\draw(14.166666666666668,53.333333333333336)node[anchor=south]{41};\draw(13.333333333333334,55.0)rectangle(15.0,56.66666666666667);\draw(14.166666666666668,56.66666666666667)node[anchor=north]{43};\draw(14.166666666666668,55.0)node[anchor=south]{42};\draw(13.333333333333334,56.66666666666667)rectangle(15.0,58.333333333333336);\draw(14.166666666666668,58.333333333333336)node[anchor=north]{44};\draw(14.166666666666668,56.66666666666667)node[anchor=south]{43};\draw(13.333333333333334,58.333333333333336)rectangle(15.0,60.0);\draw(14.166666666666668,60.0)node[anchor=north]{45};\draw(14.166666666666668,58.333333333333336)node[anchor=south]{44};\draw(13.333333333333334,60.0)rectangle(15.0,61.66666666666667);\draw(15.0,60.833333333333336)node[anchor=east]{46};\draw(14.166666666666668,60.0)node[anchor=south]{45};\draw(15.0,25.0)rectangle(16.666666666666668,26.666666666666668);\draw(15.0,25.833333333333336)node[anchor=west]{63};\draw(15.833333333333334,26.666666666666668)node[anchor=north]{19};\draw(15.0,26.666666666666668)rectangle(16.666666666666668,28.333333333333336);\draw(15.833333333333334,28.333333333333336)node[anchor=north]{20};\draw(15.833333333333334,26.666666666666668)node[anchor=south]{19};\draw(15.0,28.333333333333336)rectangle(16.666666666666668,30.0);\draw(15.833333333333334,30.0)node[anchor=north]{21};\draw(15.833333333333334,28.333333333333336)node[anchor=south]{20};\draw(15.0,30.0)rectangle(16.666666666666668,31.666666666666668);\draw(15.833333333333334,31.666666666666668)node[anchor=north]{22};\draw(15.833333333333334,30.0)node[anchor=south]{21};\draw(15.0,31.666666666666668)rectangle(16.666666666666668,33.333333333333336);\draw(15.833333333333334,33.333333333333336)node[anchor=north]{23};\draw(15.833333333333334,31.666666666666668)node[anchor=south]{22};\draw(15.0,33.333333333333336)rectangle(16.666666666666668,35.0);\draw(15.833333333333334,35.0)node[anchor=north]{24};\draw(15.833333333333334,33.333333333333336)node[anchor=south]{23};\draw(15.0,35.0)rectangle(16.666666666666668,36.66666666666667);\draw(15.833333333333334,36.66666666666667)node[anchor=north]{25};\draw(15.833333333333334,35.0)node[anchor=south]{24};\draw(15.0,36.66666666666667)rectangle(16.666666666666668,38.333333333333336);\draw(15.833333333333334,38.333333333333336)node[anchor=north]{26};\draw(15.833333333333334,36.66666666666667)node[anchor=south]{25};\draw(15.0,38.333333333333336)rectangle(16.666666666666668,40.0);\draw(15.833333333333334,40.0)node[anchor=north]{27};\draw(15.833333333333334,38.333333333333336)node[anchor=south]{26};\draw(15.0,40.0)rectangle(16.666666666666668,41.66666666666667);\draw(15.833333333333334,41.66666666666667)node[anchor=north]{28};\draw(15.833333333333334,40.0)node[anchor=south]{27};\draw(15.0,41.66666666666667)rectangle(16.666666666666668,43.333333333333336);\draw(15.0,42.5)node[anchor=west]{29};\draw(15.833333333333334,43.333333333333336)node[anchor=north]{64};\draw(15.833333333333334,41.66666666666667)node[anchor=south]{28};\draw(15.0,43.333333333333336)rectangle(16.666666666666668,45.0);\draw(15.0,44.16666666666667)node[anchor=west]{65};\draw(15.833333333333334,43.333333333333336)node[anchor=south]{64};\draw(15.0,45.0)rectangle(16.666666666666668,46.66666666666667);\draw(15.833333333333334,46.66666666666667)node[anchor=north]{55};\draw(15.833333333333334,45.0)node[anchor=south]{56};\draw(15.0,46.66666666666667)rectangle(16.666666666666668,48.333333333333336);\draw(15.833333333333334,48.333333333333336)node[anchor=north]{54};\draw(15.833333333333334,46.66666666666667)node[anchor=south]{55};\draw(15.0,48.333333333333336)rectangle(16.666666666666668,50.0);\draw(15.833333333333334,50.0)node[anchor=north]{53};\draw(15.833333333333334,48.333333333333336)node[anchor=south]{54};\draw(15.0,50.0)rectangle(16.666666666666668,51.66666666666667);\draw(15.833333333333334,51.66666666666667)node[anchor=north]{52};\draw(15.833333333333334,50.0)node[anchor=south]{53};\draw(15.0,51.66666666666667)rectangle(16.666666666666668,53.333333333333336);\draw(15.833333333333334,53.333333333333336)node[anchor=north]{51};\draw(15.833333333333334,51.66666666666667)node[anchor=south]{52};\draw(15.0,53.333333333333336)rectangle(16.666666666666668,55.0);\draw(15.833333333333334,55.0)node[anchor=north]{50};\draw(15.833333333333334,53.333333333333336)node[anchor=south]{51};\draw(15.0,55.0)rectangle(16.666666666666668,56.66666666666667);\draw(15.833333333333334,56.66666666666667)node[anchor=north]{49};\draw(15.833333333333334,55.0)node[anchor=south]{50};\draw(15.0,56.66666666666667)rectangle(16.666666666666668,58.333333333333336);\draw(15.833333333333334,58.333333333333336)node[anchor=north]{48};\draw(15.833333333333334,56.66666666666667)node[anchor=south]{49};\draw(15.0,58.333333333333336)rectangle(16.666666666666668,60.0);\draw(15.833333333333334,60.0)node[anchor=north]{47};\draw(15.833333333333334,58.333333333333336)node[anchor=south]{48};\draw(15.0,60.0)rectangle(16.666666666666668,61.66666666666667);\draw(15.0,60.833333333333336)node[anchor=west]{46};\draw(15.833333333333334,60.0)node[anchor=south]{47};\end{scope}

%% file: efficient.tex
\ignore{

{\tt{}\small{}\noindent\hbox{\phantom{}{\color{RoyalBlue}import} {\color{Green}Math}.{\color{Green}SelfAssembly}.{\color{Green}Manhattan}}

\noindent\hbox{\phantom{}{\color{RoyalBlue}import} {\color{RoyalBlue}qualified} {\color{Green}Data}.{\color{Green}Map} {\color{RoyalBlue}as} {\color{Green}M}}

\noindent\hbox{\phantom{}{\color{RoyalBlue}import} {\color{Green}Numeric}}

}}

{\tt{}\small{}\noindent\hbox{\phantom{}{\color{Purple}programme}::{\color{Green}Int}$\rightarrow${\color{Green}Program} ()}

\noindent\hbox{\phantom{}{\color{Purple}programme} n={\color{RoyalBlue}do}}

\noindent\hbox{\phantom{xx}seed 7 0}

\noindent\hbox{\phantom{xx}movey 3}

\noindent\hbox{\phantom{xx}a$\leftarrow$currentTile}

\noindent\hbox{\phantom{xx}movey n}

\noindent\hbox{\phantom{xx}a1$\leftarrow$currentTile}

\noindent\hbox{\phantom{xx}movex (-2)}

\noindent\hbox{\phantom{xx}movey 3}

\noindent\hbox{\phantom{xx}b$\leftarrow$currentTile}

\noindent\hbox{\phantom{xx}movey 1}

\noindent\hbox{\phantom{xx}c$\leftarrow$currentTile}

\noindent\hbox{\phantom{xx}movey (n-1)}

\noindent\hbox{\phantom{xx}b1$\leftarrow$currentTile}

\noindent\hbox{\phantom{xx}movex (-5)}

\noindent\hbox{\phantom{xx}gr$\leftarrow$currentTile}

\noindent\hbox{\phantom{xx}movey 2}

\noindent\hbox{\phantom{xx}gr1$\leftarrow$currentTile}

\noindent\hbox{\phantom{xx}movey 1}

\noindent\hbox{\phantom{xx}gr2$\leftarrow$currentTile}

\noindent\hbox{\phantom{xx}movey (n-1)}

\noindent\hbox{\phantom{xx}movex 1}

\noindent\hbox{\phantom{xx}movey (-n-1)}

\noindent\hbox{\phantom{xx}bot$\leftarrow$currentTile}

\noindent\hbox{\phantom{xx}movex 2}

\noindent\hbox{\phantom{xx}bind {\color{Green}N} a}

\noindent\hbox{\phantom{}}

\noindent\hbox{\phantom{xx}rewindTo a1}

\noindent\hbox{\phantom{xx}movex 1}

\noindent\hbox{\phantom{xx}movey 1}

\noindent\hbox{\phantom{xx}movex (-1)}

\noindent\hbox{\phantom{xx}bind {\color{Green}N} gr1}

\noindent\hbox{\phantom{}}

\noindent\hbox{\phantom{xx}rewindTo bot}

\noindent\hbox{\phantom{xx}rewindBy 1}

\noindent\hbox{\phantom{xx}movex 1}

\noindent\hbox{\phantom{xx}bind {\color{Green}N} c}

\noindent\hbox{\phantom{xx}rewindTo b1}

\noindent\hbox{\phantom{xx}movey 1}

\noindent\hbox{\phantom{xx}movex (-1)}

\noindent\hbox{\phantom{xx}bind {\color{Green}N} gr2}

}

%% file: general.tex
\ignore{

{\tt{}\small{}\noindent\hbox{\phantom{}{\color{RoyalBlue}import} {\color{Green}Math}.{\color{Green}SelfAssembly}.{\color{Green}Manhattan}}

\noindent\hbox{\phantom{}{\color{RoyalBlue}import} {\color{RoyalBlue}qualified} {\color{Green}Data}.{\color{Green}Map} {\color{RoyalBlue}as} {\color{Green}M}}

\noindent\hbox{\phantom{}{\color{RoyalBlue}import} {\color{Green}Numeric}}

}}

{\tt{}\small{}\noindent\hbox{\phantom{}{\color{Purple}programme}::{\color{Green}Int}$\rightarrow${\color{Green}Int}$\rightarrow${\color{Green}Program} ()}

\noindent\hbox{\phantom{}{\color{Purple}programme} n h={\color{RoyalBlue}do}}

\noindent\hbox{\phantom{xx}seed (3\^{}n+n) 0}

\noindent\hbox{\phantom{xx}{\color{Red}-- A first occurrence of the construction creates the cave.}}

\noindent\hbox{\phantom{xx}movey 2}

\noindent\hbox{\phantom{xx}a$\leftarrow$currentTile}

\noindent\hbox{\phantom{xx}movey (h-2)}

\noindent\hbox{\phantom{xx}c$\leftarrow$currentTile}

\noindent\hbox{\phantom{xx}movex (-3\^{}(n-1)-n)}

\noindent\hbox{\phantom{xx}{\color{Red}-- Start of the cave}}

\noindent\hbox{\phantom{xx}b$\leftarrow$currentTile}

\noindent\hbox{\phantom{xx}movey h}

\noindent\hbox{\phantom{xx}movex 1 {\color{Red}-- Top of the cave}}

\noindent\hbox{\phantom{xx}movey (-h+1)}

\noindent\hbox{\phantom{xx}d$\leftarrow$currentTile {\color{Red}-- Bottom of the cave}}

\noindent\hbox{\phantom{xx}{\color{Red}-- Now, move to the right, and repeat tile a}}

\noindent\hbox{\phantom{xx}movex (2*3\^{}(n-2))}

\noindent\hbox{\phantom{xx}bind {\color{Green}N} a}

\noindent\hbox{\phantom{xx}{\color{Red}-- Now, iterate n times. To avoid making the path pumpable, we need to reduce}}

\noindent\hbox{\phantom{xx}{\color{Red}-- the height (paremeter hh) each time.}}

\noindent\hbox{\phantom{xx}{\color{RoyalBlue}let} prog n hh b0 d0=}

\noindent\hbox{\phantom{xxxxxxxx}{\color{RoyalBlue}if} n$\leq$0 {\color{RoyalBlue}then}}

\noindent\hbox{\phantom{xxxxxxxxxx}return ()}

\noindent\hbox{\phantom{xxxxxxxx}{\color{RoyalBlue}else} {\color{RoyalBlue}do}}

\noindent\hbox{\phantom{xxxxxxxxxx}movey 1}

\noindent\hbox{\phantom{xxxxxxxxxx}an$\leftarrow$currentTile}

\noindent\hbox{\phantom{xxxxxxxxxx}movey (hh)}

\noindent\hbox{\phantom{xxxxxxxxxx}cn$\leftarrow$currentTile}

\noindent\hbox{\phantom{xxxxxxxxxx}movex (-3\^{}(n)+n)}

\noindent\hbox{\phantom{xxxxxxxxxx}bind {\color{Green}N} b0}

\noindent\hbox{\phantom{xxxxxxxxxx}rewindTo d0}

\noindent\hbox{\phantom{xxxxxxxxxx}movex (2*3\^{}(n-1)-n)}

\noindent\hbox{\phantom{xxxxxxxxxx}bind {\color{Green}N} an}

\noindent\hbox{\phantom{xxxxxxxxxx}b1$\leftarrow$nextTile b0}

\noindent\hbox{\phantom{xxxxxxxxxx}d1$\leftarrow$prevTile d0}

\noindent\hbox{\phantom{xxxxxxxxxx}rewindTo cn}

\noindent\hbox{\phantom{xxxxxxxxxx}prog (n-1) (hh-1) b1 d1}

\noindent\hbox{\phantom{xx}{\color{Red}-- Go back to tile c, and start iterating.}}

\noindent\hbox{\phantom{xx}rewindTo c}

\noindent\hbox{\phantom{xx}b0$\leftarrow$nextTile b}

\noindent\hbox{\phantom{xx}b1$\leftarrow$nextTile b0}

\noindent\hbox{\phantom{xx}d0$\leftarrow$prevTile d}

\noindent\hbox{\phantom{xx}prog (n-2) (h-3) b1 d0}

}\ignore{

{\tt{}\small{}\noindent\hbox{\phantom{}{\color{Purple}showf} (x,y)={\color{Brown}"("}++showFFloat ({\color{Green}Just} 4) x ({\color{Brown}","}++showFFloat ({\color{Green}Just} 4) y {\color{Brown}")"})}

\noindent\hbox{\phantom{}}

\noindent\hbox{\phantom{}{\color{Purple}main}::{\color{Green}IO} ()}

\noindent\hbox{\phantom{}{\color{Purple}main}={\color{RoyalBlue}do}}

\noindent\hbox{\phantom{xx}putStrLn {\color{Brown}"n Tileset\_size Manhattan\_distance"}}

\noindent\hbox{\phantom{xx}{\color{RoyalBlue}let} n=3}

\noindent\hbox{\phantom{xxxxxx}(seeds,tiles)=execProgram \$ programme 8 10}

\noindent\hbox{\phantom{xx}(x0,y0,x1,y1,grid) $\leftarrow$ simulate 10000 10000 seeds [] tiles}

\noindent\hbox{\phantom{xx}plot {\color{Brown}"general.pdf"}}

\noindent\hbox{\phantom{xxxx}(defaultPlot \{fontSize=0.20,showGlues={\color{Green}True},showPaths={\color{Green}False},offset=0.5,scale=10,font={\color{Brown}"Latin Modern Roman"}\}) tiles grid;}

\noindent\hbox{\phantom{}}

\noindent\hbox{\phantom{}}

\noindent\hbox{\phantom{xx}putStrLn \$ {\color{Brown}"Tileset size: "}++(show \$ {\color{Green}M}.size tiles)}

\noindent\hbox{\phantom{xx}putStrLn \$ {\color{Brown}"Maximum Manhattan distance: "}++show (x0,y0)++{\color{Brown}" to "}++show (x1,y1)}

\noindent\hbox{\phantom{xx}putStrLn \$ {\color{Brown}"("}++show ((abs \$ x1-x0)+(abs \$ y1-y0))++{\color{Brown}")"}}

}}

%% file: partially.tex
\ignore{

{\tt{}\small{}\noindent\hbox{\phantom{}{\color{RoyalBlue}import} {\color{Green}Math}.{\color{Green}SelfAssembly}.{\color{Green}Manhattan}}

\noindent\hbox{\phantom{}{\color{RoyalBlue}import} {\color{RoyalBlue}qualified} {\color{Green}Data}.{\color{Green}Map} {\color{RoyalBlue}as} {\color{Green}M}}

\noindent\hbox{\phantom{}{\color{Red}--import Prelude hiding (repeat)}}

\noindent\hbox{\phantom{}{\color{Red}--repeat=repete}}

}}

{\tt{}\small{}\noindent\hbox{\phantom{}{\color{Purple}programme}::{\color{Green}Program} ()}

\noindent\hbox{\phantom{}{\color{Purple}programme}={\color{RoyalBlue}do}}

\noindent\hbox{\phantom{xx}seed 0 0}

\noindent\hbox{\phantom{xx}movey 1}

\noindent\hbox{\phantom{xx}a0$\leftarrow$currentTile}

\noindent\hbox{\phantom{xx}{\color{Red}-- First step: grow the part that will be repeated.}}

\noindent\hbox{\phantom{xx}{\color{Red}--}}

\noindent\hbox{\phantom{xx}{\color{Red}-- Since we want to grow upwards, and then follow it downwards closely,}}

\noindent\hbox{\phantom{xx}{\color{Red}-- we need precise control over its shape.}}

\noindent\hbox{\phantom{xx}repete 15}

\noindent\hbox{\phantom{xxxx}({\color{RoyalBlue}do}}

\noindent\hbox{\phantom{xxxxxxxx}repete 20 ({\color{RoyalBlue}do} \{ movey 2;movex 1 \})}

\noindent\hbox{\phantom{xxxxxxxx}movex 1)}

\noindent\hbox{\phantom{xx}a$\leftarrow$nextTile a0}

\noindent\hbox{\phantom{xx}{\color{Red}-- Now, lay a "blocker" out, for the partially pumped paths to stop.}}

\noindent\hbox{\phantom{xx}movex 1}

\noindent\hbox{\phantom{xx}movey (-1)}

\noindent\hbox{\phantom{xx}movex (-2)}

\noindent\hbox{\phantom{xx}{\color{Red}-- Then go down.}}

\noindent\hbox{\phantom{xx}repete 15}

\noindent\hbox{\phantom{xxxx}({\color{RoyalBlue}do}}

\noindent\hbox{\phantom{xxxxxxxx}repete 20 ({\color{RoyalBlue}do} \{ movey (-2);movex (-1) \})}

\noindent\hbox{\phantom{xxxxxxxx}movex (-1))}

\noindent\hbox{\phantom{xx}{\color{Red}-- Now, build the bottom of the construction.}}

\noindent\hbox{\phantom{xx}rewindBy 6}

\noindent\hbox{\phantom{xx}c$\leftarrow$currentTile}

\noindent\hbox{\phantom{xx}eraseAfter c}

\noindent\hbox{\phantom{xx}movey (-1)}

\noindent\hbox{\phantom{}}

\noindent\hbox{\phantom{xx}{\color{RoyalBlue}let} x2=40}

\noindent\hbox{\phantom{xxxxxx}tot=21*15-1-x2}

\noindent\hbox{\phantom{xxxxxx}x0=15}

\noindent\hbox{\phantom{xxxxxx}x1=(tot-x0){\color{Bittersweet}\`{}quot\`{}}3-15}

\noindent\hbox{\phantom{}}

\noindent\hbox{\phantom{xx}{\color{Red}-- Record the three different starting tiles of exit paths.}}

\noindent\hbox{\phantom{xx}movex x0}

\noindent\hbox{\phantom{xx}start0$\leftarrow$currentTile}

\noindent\hbox{\phantom{xx}movex x1}

\noindent\hbox{\phantom{xx}start1$\leftarrow$currentTile}

\noindent\hbox{\phantom{xx}movex (tot-x1-x0)}

\noindent\hbox{\phantom{xx}start2$\leftarrow$currentTile}

\noindent\hbox{\phantom{xx}{\color{Red}-- Now, from each starting tile, grow a partially pumpable path, that will be}}

\noindent\hbox{\phantom{xx}{\color{Red}-- blocked on its way up.}}

\noindent\hbox{\phantom{}}

\noindent\hbox{\phantom{xx}{\color{Red}-- First exit path.}}

\noindent\hbox{\phantom{xx}rewindTo start2}

\noindent\hbox{\phantom{xx}pump ({\color{RoyalBlue}do}}

\noindent\hbox{\phantom{xxxxxxxxxxx}setColor blue}

\noindent\hbox{\phantom{xxxxxxxxxxx}discreteVect 16 (16*15-3))}

\noindent\hbox{\phantom{}}

\noindent\hbox{\phantom{xx}{\color{Red}-- Here, the transition to the next repetition is simple: we just move to the}}

\noindent\hbox{\phantom{xx}{\color{Red}-- right by 120 columns, lay a blocker out, so that the repeated part (from}}

\noindent\hbox{\phantom{xx}{\color{Red}-- tile a) cannot be repeated completely.}}

\noindent\hbox{\phantom{xx}rewindBy 3}

\noindent\hbox{\phantom{xx}movex 120}

\noindent\hbox{\phantom{xx}movey 2}

\noindent\hbox{\phantom{xx}movex (-1)}

\noindent\hbox{\phantom{xx}movey (-1)}

\noindent\hbox{\phantom{xx}movex (-x1-x0-5)}

\noindent\hbox{\phantom{xx}bind {\color{Green}N} a {\color{Red}-- Finally, start the repeated part again.}}

\noindent\hbox{\phantom{}}

\noindent\hbox{\phantom{}}

\noindent\hbox{\phantom{xx}{\color{Red}-- The second exit path is more complicated, since we do not want it to}}

\noindent\hbox{\phantom{xx}{\color{Red}-- collide with the first pumped path. Moreover, the discreteVect function is}}

\noindent\hbox{\phantom{xx}{\color{Red}-- used to build the most efficient vector (in terms of number of tile types)}}

\noindent\hbox{\phantom{xx}{\color{Red}-- with the given coordinates.}}

\noindent\hbox{\phantom{xx}rewindTo start1}

\noindent\hbox{\phantom{xx}pump}

\noindent\hbox{\phantom{xxxx}({\color{RoyalBlue}do}}

\noindent\hbox{\phantom{xxxxxxxx}setColor red}

\noindent\hbox{\phantom{xxxxxxxx}{\color{RoyalBlue}let} distx=tot-x0-x1+x2+1}

\noindent\hbox{\phantom{xxxxxxxxxxxx}disty=15*40}

\noindent\hbox{\phantom{xxxxxxxx}discreteVect (distx{\color{Bittersweet}\`{}quot\`{}}2) (disty{\color{Bittersweet}\`{}quot\`{}}2))}

\noindent\hbox{\phantom{xx}rewindBy 51}

\noindent\hbox{\phantom{xx}movex 10}

\noindent\hbox{\phantom{xx}repete 7}

\noindent\hbox{\phantom{xxxx}({\color{RoyalBlue}do}}

\noindent\hbox{\phantom{xxxxxxxx}repete 20 ({\color{RoyalBlue}do} \{ movey 2;movex 1 \})}

\noindent\hbox{\phantom{xxxxxxxx}movex 1)}

\noindent\hbox{\phantom{}}

\noindent\hbox{\phantom{xx}movex (x0+4)}

\noindent\hbox{\phantom{xx}movey 2}

\noindent\hbox{\phantom{xx}movex (-1)}

\noindent\hbox{\phantom{xx}movey (-1)}

\noindent\hbox{\phantom{xx}movex (-x0-4)}

\noindent\hbox{\phantom{xx}bind {\color{Green}N} a}

\noindent\hbox{\phantom{}}

\noindent\hbox{\phantom{xx}{\color{Red}-- The "final" exit path is similar, but simpler: we closely follow the}}

\noindent\hbox{\phantom{xx}{\color{Red}-- repeated part.}}

\noindent\hbox{\phantom{xx}rewindTo start0}

\noindent\hbox{\phantom{xx}movey 1}

\noindent\hbox{\phantom{xx}movex 1}

\noindent\hbox{\phantom{xx}pump}

\noindent\hbox{\phantom{xxxx}({\color{RoyalBlue}do}}

\noindent\hbox{\phantom{xxxxxxxx}setColor green}

\noindent\hbox{\phantom{xxxxxxxx}repete 149 ({\color{RoyalBlue}do} \{ movey 2; movex 1 \}))}

\noindent\hbox{\phantom{}}

\noindent\hbox{\phantom{xx}movex 2}

\noindent\hbox{\phantom{xx}repete 2}

\noindent\hbox{\phantom{xxxx}({\color{RoyalBlue}do}}

\noindent\hbox{\phantom{xxxxxxxx}repete 148 ({\color{RoyalBlue}do} \{ movey 2; movex 1 \})}

\noindent\hbox{\phantom{xxxxxxxx}movex 2)}

\noindent\hbox{\phantom{}}

\noindent\hbox{\phantom{xx}movex 10}

\noindent\hbox{\phantom{xx}movey 2}

\noindent\hbox{\phantom{xx}movex (-1)}

\noindent\hbox{\phantom{xx}movey (-1)}

\noindent\hbox{\phantom{xx}movex (-2)}

\noindent\hbox{\phantom{xx}bind {\color{Green}N} a}

\noindent\hbox{\phantom{}}

}\ignore{

{\tt{}\small{}\noindent\hbox{\phantom{}{\color{Purple}main}::{\color{Green}IO} ()}

\noindent\hbox{\phantom{}{\color{Purple}main}={\color{RoyalBlue}do}}

\noindent\hbox{\phantom{xx}l$\leftarrow$mapM ($\lambda\,$it$\rightarrow${\color{RoyalBlue}do}}

\noindent\hbox{\phantom{xxxxxxxxxxxxxx}{\color{RoyalBlue}let} (seeds,tiles)=execProgram programme}

\noindent\hbox{\phantom{xxxxxxxxxxxxxx}print seeds}

\noindent\hbox{\phantom{xxxxxxxxxxxxxx}(x0,y0,x1,y1,grid)$\leftarrow$simulate 2000 5000 seeds [] tiles}

\noindent\hbox{\phantom{xxxxxxxxxxxxxx}plot {\color{Brown}"programme.pdf"} (defaultPlot \{ showGlues={\color{Green}True},offset=10\}) tiles grid;}

\noindent\hbox{\phantom{xxxxxxxxxxxxxx}putStrLn \$ {\color{Brown}"Tileset size: "}++(show \$ {\color{Green}M}.size tiles)}

\noindent\hbox{\phantom{xxxxxxxxxxxxxx}{\color{RoyalBlue}let} x=x1-x0}

\noindent\hbox{\phantom{xxxxxxxxxxxxxxxxxx}y=y1-y0}

\noindent\hbox{\phantom{xxxxxxxxxxxxxx}putStrLn \$ {\color{Brown}"Maximum Manhattan distance: "}++(show \$ abs x+abs y)++{\color{Brown}" "}++(show (x,y))}

\noindent\hbox{\phantom{xxxxxxxxxxxxxx}return (it,abs x+abs y, {\color{Green}M}.size tiles)}

\noindent\hbox{\phantom{xxxxxxxxxx}) [3]}

\noindent\hbox{\phantom{}}

\noindent\hbox{\phantom{xx}writeFile {\color{Brown}"graph"} \$ unlines \$ map ($\lambda\,$(it,a,b)$\rightarrow$show it++{\color{Brown}" "}++show a++{\color{Brown}" "}++show b) l}

}}

%% file: eff3.tex
\draw[use as bounding box,draw=none,fill=none](0.0000,0.0000)rectangle(2.0000,5.4000);\begin{scope}[transform canvas={scale=0.12}]\draw(0.0,20.0)rectangle(1.6666666666666667,21.666666666666668);\draw(0.8333333333333334,21.666666666666668)node[anchor=north]{20};\draw(1.6666666666666667,20.833333333333336)node[anchor=east]{19};\draw(0.0,21.666666666666668)rectangle(1.6666666666666667,23.333333333333336);\draw(0.8333333333333334,23.333333333333336)node[anchor=north]{21};\draw(0.8333333333333334,21.666666666666668)node[anchor=south]{20};\draw(0.0,23.333333333333336)rectangle(1.6666666666666667,25.0);\draw(0.8333333333333334,25.0)node[anchor=north]{22};\draw(0.8333333333333334,23.333333333333336)node[anchor=south]{21};\draw(0.0,25.0)rectangle(1.6666666666666667,26.666666666666668);\draw(0.8333333333333334,26.666666666666668)node[anchor=north]{23};\draw(0.8333333333333334,25.0)node[anchor=south]{22};\draw(0.0,26.666666666666668)rectangle(1.6666666666666667,28.333333333333336);\draw(0.8333333333333334,28.333333333333336)node[anchor=north]{24};\draw(0.8333333333333334,26.666666666666668)node[anchor=south]{23};\draw(0.0,28.333333333333336)rectangle(1.6666666666666667,30.0);\draw(1.6666666666666667,29.166666666666668)node[anchor=east]{25};\draw(0.8333333333333334,28.333333333333336)node[anchor=south]{24};\draw(1.6666666666666667,20.0)rectangle(3.3333333333333335,21.666666666666668);\draw(1.6666666666666667,20.833333333333336)node[anchor=west]{19};\draw(3.3333333333333335,20.833333333333336)node[anchor=east]{18};\draw(1.6666666666666667,21.666666666666668)rectangle(3.3333333333333335,23.333333333333336);\draw(2.5,23.333333333333336)node[anchor=north]{29};\draw(3.3333333333333335,22.5)node[anchor=east]{30};\draw(1.6666666666666667,23.333333333333336)rectangle(3.3333333333333335,25.0);\draw(2.5,25.0)node[anchor=north]{28};\draw(3.3333333333333335,24.166666666666668)node[anchor=east]{35};\draw(2.5,23.333333333333336)node[anchor=south]{29};\draw(1.6666666666666667,25.0)rectangle(3.3333333333333335,26.666666666666668);\draw(2.5,26.666666666666668)node[anchor=north]{27};\draw(2.5,25.0)node[anchor=south]{28};\draw(1.6666666666666667,26.666666666666668)rectangle(3.3333333333333335,28.333333333333336);\draw(2.5,28.333333333333336)node[anchor=north]{26};\draw(2.5,26.666666666666668)node[anchor=south]{27};\draw(1.6666666666666667,28.333333333333336)rectangle(3.3333333333333335,30.0);\draw(1.6666666666666667,29.166666666666668)node[anchor=west]{25};\draw(2.5,28.333333333333336)node[anchor=south]{26};\draw(1.6666666666666667,30.0)rectangle(3.3333333333333335,31.666666666666668);\draw(2.5,31.666666666666668)node[anchor=north]{22};\draw(3.3333333333333335,30.833333333333336)node[anchor=east]{37};\draw(1.6666666666666667,31.666666666666668)rectangle(3.3333333333333335,33.333333333333336);\draw(2.5,33.333333333333336)node[anchor=north]{23};\draw(2.5,31.666666666666668)node[anchor=south]{22};\draw(1.6666666666666667,33.333333333333336)rectangle(3.3333333333333335,35.0);\draw(2.5,35.0)node[anchor=north]{24};\draw(2.5,33.333333333333336)node[anchor=south]{23};\draw(1.6666666666666667,35.0)rectangle(3.3333333333333335,36.66666666666667);\draw(3.3333333333333335,35.833333333333336)node[anchor=east]{25};\draw(2.5,35.0)node[anchor=south]{24};\draw(3.3333333333333335,20.0)rectangle(5.0,21.666666666666668);\draw(3.3333333333333335,20.833333333333336)node[anchor=west]{18};\draw(5.0,20.833333333333336)node[anchor=east]{17};\draw(3.3333333333333335,21.666666666666668)rectangle(5.0,23.333333333333336);\draw(3.3333333333333335,22.5)node[anchor=west]{30};\draw(5.0,22.5)node[anchor=east]{31};\draw(3.3333333333333335,23.333333333333336)rectangle(5.0,25.0);\draw(3.3333333333333335,24.166666666666668)node[anchor=west]{35};\draw(4.166666666666667,25.0)node[anchor=north]{12};\draw(3.3333333333333335,25.0)rectangle(5.0,26.666666666666668);\draw(4.166666666666667,26.666666666666668)node[anchor=north]{13};\draw(4.166666666666667,25.0)node[anchor=south]{12};\draw(3.3333333333333335,26.666666666666668)rectangle(5.0,28.333333333333336);\draw(4.166666666666667,28.333333333333336)node[anchor=north]{14};\draw(4.166666666666667,26.666666666666668)node[anchor=south]{13};\draw(3.3333333333333335,28.333333333333336)rectangle(5.0,30.0);\draw(3.3333333333333335,29.166666666666668)node[anchor=west]{15};\draw(4.166666666666667,30.0)node[anchor=north]{36};\draw(4.166666666666667,28.333333333333336)node[anchor=south]{14};\draw(3.3333333333333335,30.0)rectangle(5.0,31.666666666666668);\draw(3.3333333333333335,30.833333333333336)node[anchor=west]{37};\draw(4.166666666666667,30.0)node[anchor=south]{36};\draw(3.3333333333333335,31.666666666666668)rectangle(5.0,33.333333333333336);\draw(4.166666666666667,33.333333333333336)node[anchor=north]{27};\draw(4.166666666666667,31.666666666666668)node[anchor=south]{28};\draw(3.3333333333333335,33.333333333333336)rectangle(5.0,35.0);\draw(4.166666666666667,35.0)node[anchor=north]{26};\draw(4.166666666666667,33.333333333333336)node[anchor=south]{27};\draw(3.3333333333333335,35.0)rectangle(5.0,36.66666666666667);\draw(3.3333333333333335,35.833333333333336)node[anchor=west]{25};\draw(4.166666666666667,35.0)node[anchor=south]{26};\draw(5.0,20.0)rectangle(6.666666666666667,21.666666666666668);\draw(5.0,20.833333333333336)node[anchor=west]{17};\draw(6.666666666666667,20.833333333333336)node[anchor=east]{16};\draw(5.0,21.666666666666668)rectangle(6.666666666666667,23.333333333333336);\draw(5.0,22.5)node[anchor=west]{31};\draw(5.833333333333334,23.333333333333336)node[anchor=north]{3};\draw(5.0,23.333333333333336)rectangle(6.666666666666667,25.0);\draw(5.833333333333334,25.0)node[anchor=north]{4};\draw(5.833333333333334,23.333333333333336)node[anchor=south]{3};\draw(5.0,25.0)rectangle(6.666666666666667,26.666666666666668);\draw(5.833333333333334,26.666666666666668)node[anchor=north]{5};\draw(5.833333333333334,25.0)node[anchor=south]{4};\draw(5.0,26.666666666666668)rectangle(6.666666666666667,28.333333333333336);\draw(5.833333333333334,28.333333333333336)node[anchor=north]{6};\draw(5.833333333333334,26.666666666666668)node[anchor=south]{5};\draw(5.0,28.333333333333336)rectangle(6.666666666666667,30.0);\draw(5.0,29.166666666666668)node[anchor=west]{7};\draw(6.666666666666667,29.166666666666668)node[anchor=east]{32};\draw(5.833333333333334,28.333333333333336)node[anchor=south]{6};\draw(5.0,30.0)rectangle(6.666666666666667,31.666666666666668);\draw(5.833333333333334,31.666666666666668)node[anchor=north]{21};\draw(6.666666666666667,30.833333333333336)node[anchor=east]{34};\draw(5.0,31.666666666666668)rectangle(6.666666666666667,33.333333333333336);\draw(5.833333333333334,33.333333333333336)node[anchor=north]{22};\draw(5.833333333333334,31.666666666666668)node[anchor=south]{21};\draw(5.0,33.333333333333336)rectangle(6.666666666666667,35.0);\draw(5.833333333333334,35.0)node[anchor=north]{23};\draw(5.833333333333334,33.333333333333336)node[anchor=south]{22};\draw(5.0,35.0)rectangle(6.666666666666667,36.66666666666667);\draw(5.833333333333334,36.66666666666667)node[anchor=north]{24};\draw(5.833333333333334,35.0)node[anchor=south]{23};\draw(5.0,36.66666666666667)rectangle(6.666666666666667,38.333333333333336);\draw(6.666666666666667,37.5)node[anchor=east]{25};\draw(5.833333333333334,36.66666666666667)node[anchor=south]{24};\draw(6.666666666666667,20.0)rectangle(8.333333333333334,21.666666666666668);\draw(6.666666666666667,20.833333333333336)node[anchor=west]{16};\draw(8.333333333333334,20.833333333333336)node[anchor=east]{15};\draw(6.666666666666667,21.666666666666668)rectangle(8.333333333333334,23.333333333333336);\draw(7.5,23.333333333333336)node[anchor=north]{22};\draw(8.333333333333334,22.5)node[anchor=east]{37};\draw(6.666666666666667,23.333333333333336)rectangle(8.333333333333334,25.0);\draw(7.5,25.0)node[anchor=north]{23};\draw(7.5,23.333333333333336)node[anchor=south]{22};\draw(6.666666666666667,25.0)rectangle(8.333333333333334,26.666666666666668);\draw(7.5,26.666666666666668)node[anchor=north]{24};\draw(7.5,25.0)node[anchor=south]{23};\draw(6.666666666666667,26.666666666666668)rectangle(8.333333333333334,28.333333333333336);\draw(8.333333333333334,27.5)node[anchor=east]{25};\draw(7.5,26.666666666666668)node[anchor=south]{24};\draw(6.666666666666667,28.333333333333336)rectangle(8.333333333333334,30.0);\draw(6.666666666666667,29.166666666666668)node[anchor=west]{32};\draw(7.5,30.0)node[anchor=north]{33};\draw(6.666666666666667,30.0)rectangle(8.333333333333334,31.666666666666668);\draw(6.666666666666667,30.833333333333336)node[anchor=west]{34};\draw(7.5,30.0)node[anchor=south]{33};\draw(6.666666666666667,31.666666666666668)rectangle(8.333333333333334,33.333333333333336);\draw(7.5,33.333333333333336)node[anchor=north]{28};\draw(8.333333333333334,32.5)node[anchor=east]{35};\draw(7.5,31.666666666666668)node[anchor=south]{29};\draw(6.666666666666667,33.333333333333336)rectangle(8.333333333333334,35.0);\draw(7.5,35.0)node[anchor=north]{27};\draw(7.5,33.333333333333336)node[anchor=south]{28};\draw(6.666666666666667,35.0)rectangle(8.333333333333334,36.66666666666667);\draw(7.5,36.66666666666667)node[anchor=north]{26};\draw(7.5,35.0)node[anchor=south]{27};\draw(6.666666666666667,36.66666666666667)rectangle(8.333333333333334,38.333333333333336);\draw(6.666666666666667,37.5)node[anchor=west]{25};\draw(7.5,36.66666666666667)node[anchor=south]{26};\draw(6.666666666666667,38.333333333333336)rectangle(8.333333333333334,40.0);\draw(7.5,40.0)node[anchor=north]{22};\draw(8.333333333333334,39.16666666666667)node[anchor=east]{37};\draw(6.666666666666667,40.0)rectangle(8.333333333333334,41.66666666666667);\draw(7.5,41.66666666666667)node[anchor=north]{23};\draw(7.5,40.0)node[anchor=south]{22};\draw(6.666666666666667,41.66666666666667)rectangle(8.333333333333334,43.333333333333336);\draw(7.5,43.333333333333336)node[anchor=north]{24};\draw(7.5,41.66666666666667)node[anchor=south]{23};\draw(6.666666666666667,43.333333333333336)rectangle(8.333333333333334,45.0);\draw(8.333333333333334,44.16666666666667)node[anchor=east]{25};\draw(7.5,43.333333333333336)node[anchor=south]{24};\draw(8.333333333333334,10.0)rectangle(10.0,11.666666666666668);\draw(9.166666666666668,11.666666666666668)node[anchor=north]{9};\draw(10.0,10.833333333333334)node[anchor=east]{8};\draw(8.333333333333334,11.666666666666668)rectangle(10.0,13.333333333333334);\draw(9.166666666666668,13.333333333333334)node[anchor=north]{10};\draw(9.166666666666668,11.666666666666668)node[anchor=south]{9};\draw(8.333333333333334,13.333333333333334)rectangle(10.0,15.0);\draw(9.166666666666668,15.0)node[anchor=north]{11};\draw(9.166666666666668,13.333333333333334)node[anchor=south]{10};\draw(8.333333333333334,15.0)rectangle(10.0,16.666666666666668);\draw(9.166666666666668,16.666666666666668)node[anchor=north]{12};\draw(9.166666666666668,15.0)node[anchor=south]{11};\draw(8.333333333333334,16.666666666666668)rectangle(10.0,18.333333333333336);\draw(9.166666666666668,18.333333333333336)node[anchor=north]{13};\draw(9.166666666666668,16.666666666666668)node[anchor=south]{12};\draw(8.333333333333334,18.333333333333336)rectangle(10.0,20.0);\draw(9.166666666666668,20.0)node[anchor=north]{14};\draw(9.166666666666668,18.333333333333336)node[anchor=south]{13};\draw(8.333333333333334,20.0)rectangle(10.0,21.666666666666668);\draw(8.333333333333334,20.833333333333336)node[anchor=west]{15};\draw(9.166666666666668,21.666666666666668)node[anchor=north]{36};\draw(9.166666666666668,20.0)node[anchor=south]{14};\draw(8.333333333333334,21.666666666666668)rectangle(10.0,23.333333333333336);\draw(8.333333333333334,22.5)node[anchor=west]{37};\draw(9.166666666666668,21.666666666666668)node[anchor=south]{36};\draw(8.333333333333334,23.333333333333336)rectangle(10.0,25.0);\draw(9.166666666666668,25.0)node[anchor=north]{27};\draw(9.166666666666668,23.333333333333336)node[anchor=south]{28};\draw(8.333333333333334,25.0)rectangle(10.0,26.666666666666668);\draw(9.166666666666668,26.666666666666668)node[anchor=north]{26};\draw(9.166666666666668,25.0)node[anchor=south]{27};\draw(8.333333333333334,26.666666666666668)rectangle(10.0,28.333333333333336);\draw(8.333333333333334,27.5)node[anchor=west]{25};\draw(9.166666666666668,26.666666666666668)node[anchor=south]{26};\draw(8.333333333333334,31.666666666666668)rectangle(10.0,33.333333333333336);\draw(8.333333333333334,32.5)node[anchor=west]{35};\draw(9.166666666666668,33.333333333333336)node[anchor=north]{12};\draw(8.333333333333334,33.333333333333336)rectangle(10.0,35.0);\draw(9.166666666666668,35.0)node[anchor=north]{13};\draw(9.166666666666668,33.333333333333336)node[anchor=south]{12};\draw(8.333333333333334,35.0)rectangle(10.0,36.66666666666667);\draw(9.166666666666668,36.66666666666667)node[anchor=north]{14};\draw(9.166666666666668,35.0)node[anchor=south]{13};\draw(8.333333333333334,36.66666666666667)rectangle(10.0,38.333333333333336);\draw(8.333333333333334,37.5)node[anchor=west]{15};\draw(9.166666666666668,38.333333333333336)node[anchor=north]{36};\draw(9.166666666666668,36.66666666666667)node[anchor=south]{14};\draw(8.333333333333334,38.333333333333336)rectangle(10.0,40.0);\draw(8.333333333333334,39.16666666666667)node[anchor=west]{37};\draw(9.166666666666668,38.333333333333336)node[anchor=south]{36};\draw(8.333333333333334,40.0)rectangle(10.0,41.66666666666667);\draw(9.166666666666668,41.66666666666667)node[anchor=north]{27};\draw(9.166666666666668,40.0)node[anchor=south]{28};\draw(8.333333333333334,41.66666666666667)rectangle(10.0,43.333333333333336);\draw(9.166666666666668,43.333333333333336)node[anchor=north]{26};\draw(9.166666666666668,41.66666666666667)node[anchor=south]{27};\draw(8.333333333333334,43.333333333333336)rectangle(10.0,45.0);\draw(8.333333333333334,44.16666666666667)node[anchor=west]{25};\draw(9.166666666666668,43.333333333333336)node[anchor=south]{26};\draw(10.0,10.0)rectangle(11.666666666666668,11.666666666666668);\draw(10.0,10.833333333333334)node[anchor=west]{8};\draw(11.666666666666668,10.833333333333334)node[anchor=east]{7};\draw(11.666666666666668,0.0)rectangle(13.333333333333334,1.6666666666666667);\draw(12.5,1.6666666666666667)node[anchor=north]{1};\draw(11.666666666666668,1.6666666666666667)rectangle(13.333333333333334,3.3333333333333335);\draw(12.5,3.3333333333333335)node[anchor=north]{2};\draw(12.5,1.6666666666666667)node[anchor=south]{1};\draw(11.666666666666668,3.3333333333333335)rectangle(13.333333333333334,5.0);\draw(12.5,5.0)node[anchor=north]{3};\draw(12.5,3.3333333333333335)node[anchor=south]{2};\draw(11.666666666666668,5.0)rectangle(13.333333333333334,6.666666666666667);\draw(12.5,6.666666666666667)node[anchor=north]{4};\draw(12.5,5.0)node[anchor=south]{3};\draw(11.666666666666668,6.666666666666667)rectangle(13.333333333333334,8.333333333333334);\draw(12.5,8.333333333333334)node[anchor=north]{5};\draw(12.5,6.666666666666667)node[anchor=south]{4};\draw(11.666666666666668,8.333333333333334)rectangle(13.333333333333334,10.0);\draw(12.5,10.0)node[anchor=north]{6};\draw(12.5,8.333333333333334)node[anchor=south]{5};\draw(11.666666666666668,10.0)rectangle(13.333333333333334,11.666666666666668);\draw(11.666666666666668,10.833333333333334)node[anchor=west]{7};\draw(13.333333333333334,10.833333333333334)node[anchor=east]{32};\draw(12.5,10.0)node[anchor=south]{6};\draw(11.666666666666668,11.666666666666668)rectangle(13.333333333333334,13.333333333333334);\draw(12.5,13.333333333333334)node[anchor=north]{21};\draw(13.333333333333334,12.5)node[anchor=east]{34};\draw(11.666666666666668,13.333333333333334)rectangle(13.333333333333334,15.0);\draw(12.5,15.0)node[anchor=north]{22};\draw(12.5,13.333333333333334)node[anchor=south]{21};\draw(11.666666666666668,15.0)rectangle(13.333333333333334,16.666666666666668);\draw(12.5,16.666666666666668)node[anchor=north]{23};\draw(12.5,15.0)node[anchor=south]{22};\draw(11.666666666666668,16.666666666666668)rectangle(13.333333333333334,18.333333333333336);\draw(12.5,18.333333333333336)node[anchor=north]{24};\draw(12.5,16.666666666666668)node[anchor=south]{23};\draw(11.666666666666668,18.333333333333336)rectangle(13.333333333333334,20.0);\draw(13.333333333333334,19.166666666666668)node[anchor=east]{25};\draw(12.5,18.333333333333336)node[anchor=south]{24};\draw(13.333333333333334,10.0)rectangle(15.0,11.666666666666668);\draw(13.333333333333334,10.833333333333334)node[anchor=west]{32};\draw(14.166666666666668,11.666666666666668)node[anchor=north]{33};\draw(13.333333333333334,11.666666666666668)rectangle(15.0,13.333333333333334);\draw(13.333333333333334,12.5)node[anchor=west]{34};\draw(14.166666666666668,11.666666666666668)node[anchor=south]{33};\draw(13.333333333333334,13.333333333333334)rectangle(15.0,15.0);\draw(14.166666666666668,15.0)node[anchor=north]{28};\draw(15.0,14.166666666666668)node[anchor=east]{35};\draw(14.166666666666668,13.333333333333334)node[anchor=south]{29};\draw(13.333333333333334,15.0)rectangle(15.0,16.666666666666668);\draw(14.166666666666668,16.666666666666668)node[anchor=north]{27};\draw(14.166666666666668,15.0)node[anchor=south]{28};\draw(13.333333333333334,16.666666666666668)rectangle(15.0,18.333333333333336);\draw(14.166666666666668,18.333333333333336)node[anchor=north]{26};\draw(14.166666666666668,16.666666666666668)node[anchor=south]{27};\draw(13.333333333333334,18.333333333333336)rectangle(15.0,20.0);\draw(13.333333333333334,19.166666666666668)node[anchor=west]{25};\draw(14.166666666666668,18.333333333333336)node[anchor=south]{26};\draw(13.333333333333334,20.0)rectangle(15.0,21.666666666666668);\draw(14.166666666666668,21.666666666666668)node[anchor=north]{22};\draw(15.0,20.833333333333336)node[anchor=east]{37};\draw(13.333333333333334,21.666666666666668)rectangle(15.0,23.333333333333336);\draw(14.166666666666668,23.333333333333336)node[anchor=north]{23};\draw(14.166666666666668,21.666666666666668)node[anchor=south]{22};\draw(13.333333333333334,23.333333333333336)rectangle(15.0,25.0);\draw(14.166666666666668,25.0)node[anchor=north]{24};\draw(14.166666666666668,23.333333333333336)node[anchor=south]{23};\draw(13.333333333333334,25.0)rectangle(15.0,26.666666666666668);\draw(15.0,25.833333333333336)node[anchor=east]{25};\draw(14.166666666666668,25.0)node[anchor=south]{24};\draw(15.0,13.333333333333334)rectangle(16.666666666666668,15.0);\draw(15.0,14.166666666666668)node[anchor=west]{35};\draw(15.833333333333334,15.0)node[anchor=north]{12};\draw(15.0,15.0)rectangle(16.666666666666668,16.666666666666668);\draw(15.833333333333334,16.666666666666668)node[anchor=north]{13};\draw(15.833333333333334,15.0)node[anchor=south]{12};\draw(15.0,16.666666666666668)rectangle(16.666666666666668,18.333333333333336);\draw(15.833333333333334,18.333333333333336)node[anchor=north]{14};\draw(15.833333333333334,16.666666666666668)node[anchor=south]{13};\draw(15.0,18.333333333333336)rectangle(16.666666666666668,20.0);\draw(15.0,19.166666666666668)node[anchor=west]{15};\draw(15.833333333333334,20.0)node[anchor=north]{36};\draw(15.833333333333334,18.333333333333336)node[anchor=south]{14};\draw(15.0,20.0)rectangle(16.666666666666668,21.666666666666668);\draw(15.0,20.833333333333336)node[anchor=west]{37};\draw(15.833333333333334,20.0)node[anchor=south]{36};\draw(15.0,21.666666666666668)rectangle(16.666666666666668,23.333333333333336);\draw(15.833333333333334,23.333333333333336)node[anchor=north]{27};\draw(15.833333333333334,21.666666666666668)node[anchor=south]{28};\draw(15.0,23.333333333333336)rectangle(16.666666666666668,25.0);\draw(15.833333333333334,25.0)node[anchor=north]{26};\draw(15.833333333333334,23.333333333333336)node[anchor=south]{27};\draw(15.0,25.0)rectangle(16.666666666666668,26.666666666666668);\draw(15.0,25.833333333333336)node[anchor=west]{25};\draw(15.833333333333334,25.0)node[anchor=south]{26};\end{scope}

%% file: bs.tex
\clip(0.25,-1)rectangle(11.8,3);
\foreach\x in{0,...,5}{
  \begin{scope}[xshift=2*\x cm]
    \draw(0,0)--(0,1) ..controls (0,2) and (2,2)..(2,1)--(2,0)
    ..controls(2,0.5) and (1,0.5)..(1,0)
    ..controls(1,0.5) and (0,0.5)..(0,0)--cycle;
  \end{scope}
}

\foreach\x in{0,...,3}{
  \begin{scope}[xshift=4*\x cm]
    \draw(0,1)--(0,2);
  \end{scope}
}

\foreach\x in{0,...,3}{
  \begin{scope}[xshift=4*\x cm+2 cm]
    \draw[green](0,1)--(0.2,2);
  \end{scope}
}

\draw(0,0)--(0,1) ..controls (0,2) and (2,2)..(2,1)--(2,0)
..controls(2,0.5) and (1,0.5)..(1,0)
..controls(1,0.5) and (0,0.5)..(0,0)--cycle;

  \begin{scope}[y={(0.2cm,0.6cm)}]
    \foreach\x in{0,...,6}{
      \draw[blue](2*\x-1,0)--(2*\x-1,1) ..controls (2*\x-1,2) and (2*\x+1,2)..(2*\x+1,1)--(2*\x+1,0);
    }
    \foreach\x in{0,...,6}{
      \draw[blue](4*\x-3,1)--(4*\x-3,3) ..controls (4*\x-3,5) and (4*\x+1,5)..(4*\x+1,3)--(4*\x+1,1);
    }
  \end{scope}

\begin{scope}[scale=0.5,yshift= -1cm]
  \foreach\x in{0,...,11}{
    \begin{scope}[xshift=2*\x cm]
      \draw(0,0)--(0,1) ..controls (0,2) and (2,2)..(2,1)--(2,0)
      ..controls(2,0.5) and (1,0.5)..(1,0)
      ..controls(1,0.5) and (0,0.5)..(0,0)--cycle;
    \end{scope}
  }

  \begin{scope}[y={(0.4cm,0.6cm)}]
    \foreach\x in{0,...,12}{
      \draw[red](2*\x-1,0)--(2*\x-1,1) ..controls (2*\x-1,2) and (2*\x+1,2)..(2*\x+1,1)--(2*\x+1,0);
    }
    \foreach\x in{0,...,6}{
      \draw[red](4*\x-3,1)--(4*\x-3,3) ..controls (4*\x-3,5) and (4*\x+1,5)..(4*\x+1,3)--(4*\x+1,1);
    }
  \end{scope}

  \begin{scope}[scale=0.5,yshift= -1cm]
    \foreach\x in{0,...,23}{
      \begin{scope}[xshift=2*\x cm]
        \draw(0,0)--(0,1) ..controls (0,2) and (2,2)..(2,1)--(2,0)
        ..controls(2,0.5) and (1,0.5)..(1,0)
        ..controls(1,0.5) and (0,0.5)..(0,0)--cycle;
      \end{scope}
    }
  \end{scope}
\end{scope}

%% file: bitsel.tex
\begin{scope}[draw=gray]
\foreach\x in{0,...,2}{
  \begin{scope}[yshift=2*\x cm]
    \draw(1,0)--(0,0) ..controls (-1,0) and (-1,2)..(0,2)--(1,2)
    ..controls(0.5,2) and (0.5,1)..(1,1)
    ..controls(0.5,1) and (0.5,0)..(1,0)--cycle;
  \end{scope}
}
\begin{scope}[xshift=1cm,scale=0.5]
\foreach\x in{0,...,5}{
  \begin{scope}[yshift=2*\x cm]
    \draw(1,0)--(0,0) ..controls (-1,0) and (-1,2)..(0,2)--(1,2)
    ..controls(0.5,2) and (0.5,1)..(1,1)
    ..controls(0.5,1) and (0.5,0)..(1,0)--cycle;
  \end{scope}
}
\begin{scope}[xshift=1cm,scale=0.5]
\foreach\x in{0,...,11}{
  \begin{scope}[yshift=2*\x cm]
    \draw(1,0)--(0,0) ..controls (-1,0) and (-1,2)..(0,2)--(1,2)
    ..controls(0.5,2) and (0.5,1)..(1,1)
    ..controls(0.5,1) and (0.5,0)..(1,0)--cycle;
  \end{scope}
}
\begin{scope}[xshift=1cm,scale=0.5]
\foreach\x in{0,...,23}{
  \begin{scope}[yshift=2*\x cm]
    \draw(1,0)--(0,0) ..controls (-1,0) and (-1,2)..(0,2)--(1,2)
    ..controls(0.5,2) and (0.5,1)..(1,1)
    ..controls(0.5,1) and (0.5,0)..(1,0)--cycle;
  \end{scope}
}
\end{scope}
\end{scope}
\end{scope}
\end{scope}

%% file: bitsel1.tex
\begin{scope}[scale=0.5,xshift=1cm,thick]
\draw[red,->](-2,0)--(0.5,0);
\draw[red](0.5,0)--(1,0);

\draw[red](1,0)..controls (0,0) and (0,2)..(1,2);
\draw[green](1,2)--(0,2)..controls (-2,2) and (-2,6)..(0,6)--(1,6);
\draw[red,->](1,6)--(1.5,6);
\draw[red](1.5,6)--(2,6)..controls (1.5,6) and (1.5,5)..(2,5);

\draw[green,->](2,5)--(1.5,5);
\draw[green](1.5,5)--(1,5)..controls(0,5) and (0,3)..(1,3)--(2,3);
\draw[red,->](2,3)..controls(1.5,3) and (1.5,2)..(2,2)
..controls(1.5,2) and (1.5,1)..(2,1)
..controls(1.5,1) and (1.5,0)..(2,0)--(2.5,0)--(2.75,0)--(3,0);

\end{scope}

%% file: bitsel2.tex
\begin{scope}[scale=0.5,xshift=1cm,thick]
\draw[red,->](-2,0)--(0.5,0);
\draw[red](0.5,0)--(1,0);

\draw[red,->](1,0)..controls (0,0) and (0,2)..(1,2)..controls(0,2)and(0,4)..(1,4)--(1.5,4);
\draw[red,->](1.5,4)--(2,4)..controls(1.5,4) and (1.5,3)..
(2,3)..controls(1.5,3) and (1.5,2)..(2,2)
..controls(1.5,2) and (1.5,1)..(2,1)
..controls(1.5,1) and (1.5,0)..(2,0)--(2.5,0)--(2.75,0)--(3,0);
\end{scope}